\documentclass[twocolumn]{aastex62}
\usepackage{etoolbox}

\hypersetup{linkcolor=blue,citecolor=blue,urlcolor=blue}

\accepted{to ApJS on October 25, 2019}
\shorttitle{MASSES Full Data Release}
\shortauthors{Stephens et al.}
\begin{document}
\newcommand{\ntdp}{\mbox{N$_2$D$^+$(3--2)}}
\newcommand{\ceo}{\mbox{C$^{18}$O(2--1)}}
\newcommand{\ceooz}{\mbox{C$^{18}$O(1--0)}}
\newcommand{\ttco}{\mbox{$^{13}$CO(2--1)}}
\newcommand{\coto}{\mbox{CO(2--1)}}

\newcommand{\ntdpnt}{\mbox{N$_2$D$^+$}}
\newcommand{\ceont}{\mbox{C$^{18}$O}}
\newcommand{\ttcont}{\mbox{$^{13}$CO}}
\newcommand{\cotont}{\mbox{CO}}

\newcommand{\httcop}{\mbox{H$^{13}$CO$^+$(4--3)}}
\newcommand{\hcop}{\mbox{HCO$^+$(4--3)}}
\newcommand{\cott}{\mbox{CO(3--2)}}

\newcommand{\httcopnt}{\mbox{H$^{13}$CO$^+$}}
\newcommand{\hcopnt}{\mbox{HCO$^+$}}
\newcommand{\cottnt}{\mbox{CO}}

\newcommand{\kms}{km\,s$^{-1}$}

\defcitealias{Stephens2018}{Paper~I}

\title{Mass Assembly of Stellar Systems and their Evolution with the SMA (MASSES) -- Full Data Release}

\author{Ian W. Stephens}
%\affiliation{Harvard-Smithsonian Center for Astrophysics, 60 Garden Street, Cambridge, MA, USA}
\affiliation{Center for Astrophysics $|$ Harvard \& Smithsonian, 60 Garden Street, Cambridge, MA 02138, USA \href{mailto:ian.stephens@cfa.harvard.edu}{ian.stephens@cfa.harvard.edu}} %; \href{mailto:ian.stephens@cfa.harvard.edu}{ian.stephens@cfa.harvard.edu} }

\author{Tyler L. Bourke}
\affiliation{SKA Organization, Jodrell Bank, Lower Withington, Macclesfield, Cheshire SK11 9FT, UK}
\affiliation{Center for Astrophysics $|$ Harvard \& Smithsonian, 60 Garden Street, Cambridge, MA 02138, USA \href{mailto:ian.stephens@cfa.harvard.edu}{ian.stephens@cfa.harvard.edu}} 

\author{Michael M. Dunham}
\affiliation{Department of Physics, State University of New York at Fredonia, 280 Central Ave, Fredonia, NY 14063, USA}
\affiliation{Center for Astrophysics $|$ Harvard \& Smithsonian, 60 Garden Street, Cambridge, MA 02138, USA \href{mailto:ian.stephens@cfa.harvard.edu}{ian.stephens@cfa.harvard.edu}} 

\author{Philip C. Myers}
\affiliation{Center for Astrophysics $|$ Harvard \& Smithsonian, 60 Garden Street, Cambridge, MA 02138, USA \href{mailto:ian.stephens@cfa.harvard.edu}{ian.stephens@cfa.harvard.edu}} 

\author{Riwaj Pokhrel}
\affiliation{Department of Astronomy, University of Massachusetts, Amherst, MA 01003, USA}
%\affiliation{Center for Astrophysics $|$ Harvard \& Smithsonian, 60 Garden Street, Cambridge, MA 02138, USA \href{mailto:ian.stephens@cfa.harvard.edu}{ian.stephens@cfa.harvard.edu}} 

\author{John J. Tobin}
\affiliation{National Radio Astronomy Observatory, 520 Edgemont Rd., Charlottesville, VA 22903, USA}

\author{H\'{e}ctor G. Arce}
\affiliation{Department of Astronomy, Yale University, New Haven, CT 06520, USA}

\author{Sarah I. Sadavoy}
\affiliation{Center for Astrophysics $|$ Harvard \& Smithsonian, 60 Garden Street, Cambridge, MA 02138, USA \href{mailto:ian.stephens@cfa.harvard.edu}{ian.stephens@cfa.harvard.edu}} 

%\affiliation{Homer L. Dodge Department of Physics and Astronomy, University of Oklahoma, 440 W. Brooks Street, Norman, OK 73019, USA}
%\affiliation{Leiden Observatory, Leiden University, P.O. Box 9513, 2300-RA Leiden, The Netherlands}

\author{Eduard I. Vorobyov}
\affiliation{Research Institute of Physics, Southern Federal University, Stachki Ave. 194, Rostov-on-Don, 344090, Russia}
\affiliation{University of Vienna, Department of Astrophysics, Vienna, A-1180, Austria}

\author{Jaime E. Pineda}
\affiliation{Max-Planck-Institut f\"ur extraterrestrische Physik, D-85748 Garching, Germany}

\author{Stella S. R. Offner}
\affiliation{Department of Astronomy, The University of Texas at Austin, Austin, TX 78712, USA}

\author{Katherine I. Lee}
\affiliation{Center for Astrophysics $|$ Harvard \& Smithsonian, 60 Garden Street, Cambridge, MA 02138, USA \href{mailto:ian.stephens@cfa.harvard.edu}{ian.stephens@cfa.harvard.edu}} 

\author{Lars E. Kristensen}
\affiliation{Niels Bohr Institute and Center for Star and Planet Formation, Copenhagen University, DK-1350 Copenhagen K., Denmark}
%\affiliation{Centre for Star and Planet Formation, Niels Bohr Institute and Natural History Museum of Denmark, University of Copenhagen, \O ster Voldgade 5-7, DK-1350 Copenhagen K, Denmark}

\author{Jes K. J\o rgensen}
\affiliation{Niels Bohr Institute and Center for Star and Planet Formation, Copenhagen University, DK-1350 Copenhagen K., Denmark}

\author{Mark A. Gurwell}
\affiliation{Center for Astrophysics $|$ Harvard \& Smithsonian, 60 Garden Street, Cambridge, MA 02138, USA \href{mailto:ian.stephens@cfa.harvard.edu}{ian.stephens@cfa.harvard.edu}} 

\author{Alyssa A. Goodman}
\affiliation{Center for Astrophysics $|$ Harvard \& Smithsonian, 60 Garden Street, Cambridge, MA 02138, USA \href{mailto:ian.stephens@cfa.harvard.edu}{ian.stephens@cfa.harvard.edu}}



\begin{abstract}
We present and release the full dataset for the Mass Assembly of Stellar Systems and their Evolution with the SMA (MASSES) survey. This survey used the Submillimeter Array (SMA) to image the 74 known protostars within the Perseus molecular cloud. The SMA was used in two array configurations to capture outflows for scales $>$30$\arcsec$ ($>$9000\,au) and to probe scales down to $\sim$1$\arcsec$ ($\sim$300\,au). The protostars were observed with the 1.3\,mm and 850\,$\mu$m receivers simultaneously to detect continuum at both wavelengths and molecular line emission from \coto, \ttco, \ceo, \ntdp, \cott, \hcop, and \httcop. Some of the observations also used the SMA's recently upgraded correlator, SWARM, whose broader bandwidth allowed for several more spectral lines to be observed (e.g., SO, H$_2$CO, DCO$^+$, DCN, CS, CN). Of the main continuum and spectral tracers observed, 84\% of the images and cubes had emission detected. The median \ceo\ linewidth is $\sim$1.0\,\kms, which is slightly higher than those measured with single-dish telescopes at scales of 3000--20,000\,au.  Of the 74 targets, six are suggested to be first hydrostatic core candidates, and we suggest that L1451-mm is the best candidate. We question a previous continuum detection toward L1448~IRS2E. In the SVS~13 system, SVS~13A certainly appears to be the most evolved source, while SVS~13C appears to be hotter and more evolved than SVS~13B. The MASSES survey is the largest publicly available interferometric continuum and spectral line protostellar survey to date, and is largely unbiased as it only targets protostars in Perseus. All visibility ($uv$) data and imaged data are publicly available at \url{https://dataverse.harvard.edu/dataverse/full_MASSES/}.

%Some of the observations also used the SMA's recently upgraded correlator, SWARM; its broader bandwidth allowed for several more spectral lines to be covered (e.g., SO, H$_2$CO, DCO$^+$, DCN, CS, CN). We measure typical protostellar \ceo\ linewidths of 1.0\,\kms, which is slightly higher than that measured with single-dish telescopes at scales of 3000 to 20000\,au. We find that 6 sources originally identified as Class~0 or~I protostars may not actually be protostars. Of the 74 targets, six are suggested to be first hydrostatic core candidates, and we suggest that L1451-mm is the best candidates. We question a previous continuum detection toward L1448~IRS2E.  Over 30 spectral lines are detected toward the SVS~13 system with SWARM, and we find substantial variation of deuterated species between the three main protostars. SVS~13A certainly appears to be the most evolved source, while SVS~13C appears to be hotter and more evolved than SVS~13B. The data presented in this paper improves on and supersedes those presented in Stephens et al. (2018), especially since some corrections to the data calibration and imaging have been implemented. All visibility ($uv$) data and imaged data are publicly available at {\bf LINK}.

\end{abstract}



\section{Introduction} \label{sec:intro}
%Talk about extra SWARM for Per44 345 ghz

The formation of stars involves a complex interaction of related phenomena. Gravitational collapse occurs within dense regions of gas and dust with magnetic fields, turbulence, stellar feedback, and rotation. Given the complexity, the balance of these ingredients may differ significantly from source to source. As such, a statistical approach is necessary to understand the typical physical and kinematic conditions of the star formation process at the early stages. For the formation of a low-mass star, the protostellar stages are often separated into two classes \citep[e.g.,][]{Andre2000,Enoch2009,Evans2009}. The youngest is the Class~0 phase, which is followed by the Class~I phase where the envelope is typically less massive. The lifetimes of these phases have considerable uncertainties, with current estimates suggesting the Class~0 and~I phase together last $\sim$0.5\,Myr \citep[e.g.,][]{Dunham2014a}. 

At scales of $\sim$500\,au--10,000\,au, protostars are enshrouded by dense envelopes that feed the protostar--disk systems, and much of this gas is ejected via outflows back into the interstellar medium. \citet{Arce2006} conducted one of the first interferometric molecular line surveys of protostars.  From a sample of roughly a dozen systems, they found evidence that outflows are more collimated at earlier stages and that the surrounding envelopes become swept up with the outflow. The outflows widen with time, carving out a wide cavity, and the envelope starts to dissipate and lengthen perpendicular to the outflow. This picture is supported by simulations \citep[e.g.,][]{Offner2011a}. Several other molecular line interferometric surveys at these scales have also provided constraints on protostellar evolution, accretion, and angular momentum transfer \citep[e.g.,][]{Jorgensen2007,Jorgensen2009,Yen2015}. However, all these studies only looked at one or two dozen protostars in various clouds, selecting primarily the brightest known protostars. An optimal sample would target all protostars within a single molecular cloud because such a sample would not be biased toward the brightest protostars, and each protostar within a single cloud is more likely to have similar environmental conditions.

For clouds within the distance of Orion \citep[$\sim$430\,pc;][]{Zucker2019}, the Perseus molecular cloud \citep[$\sim$300~pc;][]{Zucker2018,Zucker2019} has the largest number of Class~0 and~I protostars \citep{Dunham2015} and therefore is an excellent laboratory for studying the earliest stages of protostellar evolution. Moreover, the underlying multiplicity of the protostars are well known, as the Very Large Array (VLA) Nascent Disk and Multiplicity (VANDAM) survey observed every protostar at very high ($\sim$20\,au) resolution \citep{Tobin2016}. Thus, we embarked on a Submillimeter Array \citep[SMA;][]{Ho2004} large program called the Mass Assembly of Stellar Systems and their Evolution with the SMA (MASSES), where we observed molecular line and continuum emission about every protostar within Perseus. By targeting all the known protostars within a single cloud, the survey is largely unbiased. Our primary goals are to better understand the evolution, kinematics, fragmentation, and chemistry of protostars and their surrounding envelope via an unbiased statistical analysis. Thus, we targeted molecular lines that primarily trace the protostellar envelopes and outflows. Specifically, we chose \coto, \cott, and \ttco\ to trace the outflows; continuum, \ceo, \ntdp, and \httcop\ to trace the envelopes; and \hcop\ to trace both.

To cover both large scales and small scales, we used the SMA's most compact configuration (called ``subcompact" or SUB) and SMA's extended configuration (EXT). The SUB data were made publicly available in \citet[][henceforth, Paper~I]{Stephens2018}. This paper serves as a data release of the entire MASSES survey, and given the improvements and fixes mentioned in this paper (see Sections~\ref{sec:datacal} and~\ref{sec:imaging}), {\bf it is strongly advised to use these data products over those in \citetalias{Stephens2018}.}

The MASSES survey has already had considerable success for providing statistical constraints of the star formation process. In \citet{Lee2016} and \citet{Stephens2017b}, we found that the angular momentum axes of the Perseus protostars are independent of its nearest companion or the cloud's filamentary structure. These results were found to be consistent with the predictions of turbulent fragmentation \citep{Offner2016}. We found that, based on the measured sizes of the \ceo\ envelopes and predicted freeze-out timescales, protostars may experience episodes of strong accretion bursts \citep{Frimann2017}. We also showed that the Perseus molecular cloud fragments hierarchically into multiple substructures, and the fragmentation at each scale could be explained by inefficient thermal Jeans fragmentation \citep{Pokhrel2018}. Finally, we found that disk masses can be accurately constrained even with a beam size that is $\sim$10 times larger than the disk \citep{Andersen2019}, and we have used the MASSES survey to constrain the physical properties of individual protostars \citep{Lee2015,AgurtoGangas2019}.

\section{Target Selection and Correlator Setup}\label{sec:setup}
The source selection criteria are described in detail in \citetalias{Stephens2018}. Specifically, we target all the known Class~0 and~I protostars in the Perseus molecular cloud (most of which were originally classified with $Spitzer$ in \citealt{Enoch2009}), along with first hydrostatic core candidates. For the rest of this paper, we will collectively refer to all these sources as ``protostars" and refer to all targets of the MASSES survey as ``MASSES protostars." 
Table~\ref{tab:sources} lists the protostellar targets and their bolometric temperature $T_{\text{bol}}$, which is defined as the temperature of a blackbody with the same flux-weighted mean frequency, and is calculated using the observed spectral energy distribution of each protostar \citep{Myers1993}.  $T_{\text{bol}}$ is expected to be approximately correlated with the age of the protostar \citep[e.g.,][]{Young2005,Dunham2010a}. The Per-emb nomenclature for the protostars come from \citet{Enoch2009}, and the numbered suffixes were ordered by increasing $T_{\text{bol}}$. However, since then, several of the $T_{\text{bol}}$ values have been updated to more accurate values \citep[e.g.,][]{Tobin2016}, so the Per-emb names are no longer in perfect order with their bolometric temperatures.

MASSES targeted a total of 74 protostellar candidates in 68 total pointings, as we could sometimes image two protostars with a single pointing. Note that we technically observed more than 74 protostellar candidates because many of the protostars were identified by $Spitzer$, and these protostars can be resolved into multiple systems at higher resolution \citep[e.g., in the VANDAM survey;][]{Tobin2016}.

 \startlongtable
\begin{deluxetable*}{cccccccccccccccccccc}
%\tablecolumns{20}
\tabletypesize{\scriptsize}
%\tablewidth{-20pt}
\tablecaption{Source and Observing Information \label{tab:sources}}
\tablehead{\colhead{Source} & \colhead{$T_{\text{bol}}$\tablenotemark{a}} & \colhead{Other Names\tablenotemark{b}}  & \colhead{R.A.\tablenotemark{c}} & \colhead{Decl.\tablenotemark{c}} & \colhead{SMA Track} & \colhead{Array\tablenotemark{d}} & \colhead{Missing} & \colhead{Correlator} \vspace{-8pt} \\
\colhead{Name} & (K) & & \colhead{(J2000)} & \colhead{(J2000)} & Name & Config. & \colhead{Antennas\tablenotemark{e}} &  \colhead{for Track}}
\startdata
Per-emb-1 & 27 $\pm$ 1 & HH 211-MMS & 03:43:56.53 & 32:00:52.90 & 141207\_05:11:03 & SUB & 6 & ASIC \\
-- & -- & -- & -- & -- & 150218\_03:46:42\tablenotemark{f} & EXT & 6, 7 & ASIC \\
Per-emb-2 & 27 $\pm$ 1 & IRAS 03292+3039 & 03:32:17.95 & 30:49:47.60 & 141122\_03:05:36 & SUB & 6 & ASIC \\
-- & -- & -- & -- & -- & 140906\_06:18:15 & EXT & 5 & ASIC \\
Per-emb-3 & 32 $\pm$ 2 & ... & 03:29:00.52 & 31:12:00.70 & 151022\_10:48:26 & SUB & 5, 7 & ASIC \\
-- & -- & -- & -- & -- & 180918\_08:32:35 & SUB & 6 & SWARM \\
-- & -- & -- & -- & -- & 161007\_05:58:24 & EXT & 2 & SWARM \\
-- & -- & -- & -- & -- & 161023\_07:37:06 & EXT & 2 & SWARM \\
Per-emb-4 & 31 $\pm$ 3 & ... & 03:28:39.10 & 31:06:01.80 & 151102\_04:48:11 & SUB & 7 & ASIC \\
-- & -- & -- & -- & -- & 161113\_03:41:32 & EXT & none & SWARM \\
Per-emb-5 & 32 $\pm$ 2 & IRAS 03282+3035 & 03:31:20.96 & 30:45:30.205 & 141122\_03:05:36 & SUB & 6 & ASIC \\
-- & -- & -- & -- & -- & 140906\_06:18:15 & EXT & 5 & ASIC \\
Per-emb-6 & 52 $\pm$ 3 & ... & 03:33:14.40 & 31:07:10.90 & 180911\_09:12:02 & SUB & 2 & SWARM \\
%& 52 $\pm$ 3 & ... & 03:33:14.40 & 31:07:10.90 & 151203\_05:02:22 & SUB & none & ASIC \\
-- & -- & -- & -- & -- & 161007\_05:58:24 & EXT & 2 & SWARM \\
-- & -- & -- & -- & -- & 161023\_07:37:06 & EXT & 2 & SWARM \\
Per-emb-7 & 37 $\pm$ 4 & ... & 03:30:32.68 & 30:26:26.50 & 160925\_08:16:53\tablenotemark{g} & SUB & 2 & SWARM \\
-- & -- & -- & -- & -- & 160208\_05:18:40 & COM & 4, 5, 7 & ASIC \\
-- & -- & -- & -- & -- & 160209\_03:40:33 & COM & 4, 5, 7 & ASIC \\
-- & -- & -- & -- & -- & 161104\_04:21:48& EXT & 2 & SWARM \\
Per-emb-8 & 43 $\pm$ 6 & ... & 03:44:43.62 & 32:01:33.70 & 151123\_03:56:56\tablenotemark{h} & SUB & none & ASIC \\
-- & -- & -- & -- & -- & 151130\_04:08:59\tablenotemark{h} & SUB & none & ASIC \\
-- & -- & -- & -- & -- & 161023\_07:37:06 & EXT & 2 & SWARM \\
Per-emb-9 & 36 $\pm$ 2 & IRAS 03267+3128, Perseus~5 & 03:29:51.82 & 31:39:06.10 & 151023\_11:04:02 & SUB & 5, 7 & ASIC \\
-- & -- & -- & -- & -- & 151023\_14:42:17 & SUB & 5, 7 & ASIC \\
-- & -- & -- & -- & -- & 151024\_11:25:32 & SUB & 7, 8 & ASIC \\
-- & -- & -- & -- & -- & 180918\_08:32:35 & SUB & 6 & SWARM \\
-- & -- & -- & -- & -- & 161024\_04:36:52 & EXT & 2 & SWARM \\
Per-emb-10  & 30 $\pm$ 2 & ... & 03:33:16.45 & 31:06:52.50 & 180911\_09:12:02 & SUB & 2 & SWARM \\
%& 30 $\pm$ 2 & ... & 03:33:16.45 & 31:06:52.50 & 151203\_05:02:22 & SUB & none & ASIC \\
-- & -- & -- & -- & -- & 161009\_06:02:18 & EXT & 2 & SWARM \\
Per-emb-11 & 30 $\pm$ 2 & IC~348MMS & 03:43:56.85 & 32:03:04.60 & 141207\_05:11:03 & SUB & 6 & ASIC \\
-- & -- & -- & -- & -- & 150914\_10:02:49 & EXT & 7 & ASIC \\
-- & -- & -- & -- & -- & 150914\_13:02:16 & EXT & 7 & ASIC \\
-- & -- & -- & -- & -- & 161009\_06:02:18 & EXT & 2 & SWARM \\
Per-emb-12 & 29 $\pm$ 2 & NGC 1333 IRAS~4A & 03:29:10.50 & 31:13:31.00 & 141123\_04:09:39 & SUB & 6, 7, 8 & ASIC \\
-- & -- & -- & -- & -- & 141123\_07:49:31 & SUB & 6, 7, 8 & ASIC \\
-- & -- & -- & -- & -- & 141213\_03:41:25 & SUB & 6 & ASIC \\
-- & -- & -- & -- & -- & 140907\_06:16:26  & EXT & 5, 6 & ASIC \\
Per-emb-13 & 28 $\pm$ 1 & NGC 1333 IRAS~4B & 03:29:12.04 & 31:13:01.50 & 141120\_03:58:22 & SUB & 6 & ASIC \\
-- & -- & -- & -- & -- & 140904\_07:20:08 & EXT & none & ASIC \\
Per-emb-14 & 31 $\pm$ 2 & NGC 1333 IRAS~4C & 03:29:13.52 & 31:13:58.00 & 141123\_04:09:39 & SUB & 6, 7, 8 & ASIC \\
-- & -- & -- & -- & -- & 141123\_07:49:31 & SUB & 6, 7, 8 & ASIC \\
-- & -- & -- & -- & -- & 141213\_03:41:25 & SUB & 6 & ASIC \\
-- & -- & -- & -- & -- & 140907\_06:16:26  & EXT & 5, 6 & ASIC \\
Per-emb-15 & 36 $\pm$ 4 & RNO~15-FIR & 03:29:04.05 & 31:14:46.60 & 151023\_11:04:02 & SUB & 5, 7 & ASIC \\
-- & -- & -- & -- & -- & 151023\_14:42:17 & SUB & 5, 7 & ASIC \\
-- & -- & -- & -- & -- & 151024\_11:25:32 & SUB & 7, 8 & ASIC \\
-- & -- & -- & -- & -- & 160925\_08:16:53\tablenotemark{g} & SUB & 2 & SWARM \\
-- & -- & -- & -- & -- & 161031\_04:22:41 & EXT & 2 & SWARM \\
Per-emb-16 & 39 $\pm$ 2 & ... & 03:43:50.96 & 32:03:16.70 & 141207\_05:11:03 & SUB & 6 & ASIC \\
-- & -- & -- & -- & -- & 150914\_10:02:49 & EXT & 7 & ASIC \\
-- & -- & -- & -- & -- & 150914\_13:02:16 & EXT & 7 & ASIC \\
-- & -- & -- & -- & -- & 161010\_05:49:55 & EXT & 2 & SWARM \\
Per-emb-17 & 59 $\pm$ 11 & ... & 03:27:39.09 & 30:13:03.00& 151102\_04:48:11 & SUB & 7 & ASIC \\
-- & -- & -- & -- & -- & 161010\_05:49:55 & EXT & 2 & SWARM \\
Per-emb-18 & 59 $\pm$ 12 & NGC 1333 IRAS~7 & 03:29:10.99 & 31:18:25.50 & 141127\_02:21:26 & SUB & 6 & ASIC \\
-- & -- & -- & -- & -- & 150915\_10:07:22 & EXT & 7 & ASIC \\
-- & -- & -- & -- & -- & 161024\_04:36:52 & EXT & 2 & SWARM \\
Per-emb-19 & 60 $\pm$ 3 & ... & 03:29:23.49 & 31:33:29.50 & 141214\_03:50:32 & SUB & 6 & ASIC \\
-- & -- & -- & -- & -- & 151006\_05:42:22 & EXT & 7 & both \\
Per-emb-20 & 65 $\pm$ 3 & L1455-IRS4 & 03:27:43.23 & 30:12:28.80& 151108\_04:20:52 & SUB & none & ASIC \\
-- & -- & -- & -- & -- & 161011\_05:52:03 & EXT & 2 & SWARM \\
Per-emb-21 & 45 $\pm$ 12 & ... & \multicolumn{6}{c}{Imaged in the same pointing as Per-emb-18} \\
Per-emb-22 & 43 $\pm$ 2 & L1448IRS2 & 03:25:22.33 & 30:45:14.00 & 141129\_03:04:09 & SUB & 6 & ASIC \\
-- & -- & -- & -- & -- & 150228\_05:34:15 & EXT & 5, 7 & ASIC \\
-- & -- & -- & -- & -- & 150922\_06:30:31 & EXT & 7, 8 & both \\
Per-emb-23 & 42 $\pm$ 2 & ASR 30 & 03:29:17.16 & 31:27:46.40 & 151206\_04:31:17\tablenotemark{h} & SUB & none & ASIC \\
-- & -- & -- & -- & -- & 161011\_05:52:03 & EXT & 2 & SWARM \\
Per-emb-24 & 67 $\pm$ 10 & ... & 03:28:45.30 & 31:05:42.00 & 151122\_11:23:42\tablenotemark{h} & SUB & none & ASIC \\
-- & -- & -- & -- & -- & 151122\_12:21:59\tablenotemark{h} & SUB & none & ASIC \\
-- & -- & -- & -- & -- & 151127\_04:06:10\tablenotemark{h} & SUB & none & ASIC \\
-- & -- & -- & -- & -- & 161024\_04:36:52 & EXT & 2 & SWARM \\
Per-emb-25 & 61 $\pm$ 12 & ... & 03:26:37.46 & 30:15:28.00 & 151026\_05:33:00 & SUB & 7, 8 & ASIC \\
-- & -- & -- & -- & -- & 161027\_04:45:48 & EXT & 2, 6 & SWARM \\
Per-emb-26 & 47 $\pm$ 7 & L1448C, L1448-mm & 03:25:38.95 & 30:44:02.00 & 141118\_02:15:14 & SUB & 6 & ASIC \\
-- & -- & -- & -- & -- & 140905\_07:31:26 & EXT & none & ASIC \\
Per-emb-27 & 69 $\pm$ 1 & NGC 1333 IRAS~2A & 03:28:55.56 & 31:14:36.60 & 141120\_03:58:22 & SUB & 6 & ASIC \\
-- & -- & -- & -- & -- & 140904\_07:20:08 & EXT & none & ASIC \\
Per-emb-28 & 45 $\pm$ 2 & ... & \multicolumn{6}{c}{Imaged in the same pointing as Per-emb-16} \\
Per-emb-29 & 48 $\pm$ 1 & B1-c & 03:33:17.85 & 31:09:32.00 & 141128\_03:49:43 & SUB & 6 & ASIC \\
-- & -- & -- & -- & -- & 150228\_05:34:15 & EXT & 5, 7 & ASIC \\
-- & -- & -- & -- & -- & 150922\_06:30:31 & EXT & 7, 8 & both \\
Per-emb-30 & 78 $\pm$ 6 & ... & 03:33:27.28 & 31:07:10.20 & 160917\_08:50:40\tablenotemark{g} & SUB & 2 & SWARM \\
-- & -- & -- & -- & -- & 160927\_08:02:56\tablenotemark{g} & SUB & 2, 3, 6 & SWARM \\
-- & -- & -- & -- & -- & 170122\_03:03:39 & SUB & 3 & SWARM \\
-- & -- & -- & -- & -- & 170122\_14:18:47 & SUB & 3 & SWARM \\
-- & -- & -- & -- & -- & 160206\_03:00:59 & COM & 5, 7 & ASIC \\
-- & -- & -- & -- & -- & 161030\_04:24:22 & EXT & 2 & SWARM \\
Per-emb-31 & 80 $\pm$ 13  & ... & 03:28:32.55 & 31:11:05.20& 151108\_04:20:52 & SUB & none & ASIC \\
-- & -- & -- & -- & -- & 161027\_04:45:48 & EXT & 2, 6 & SWARM \\
Per-emb-32 & 57 $\pm$ 10 & ... & 03:44:02.40 & 32:02:04.90 & 151123\_03:56:56\tablenotemark{h} & SUB & none & ASIC \\
-- & -- & -- & -- & -- & 151130\_04:08:59\tablenotemark{h} & SUB & none & ASIC \\
-- & -- & -- & -- & -- & 161027\_04:45:48 & EXT & 2, 6 & SWARM \\
Per-emb-33 & 57 $\pm$ 3 & L1448IRS3B, L1448N & 03:25:36.48 & 30:45:22.30 & 141118\_02:15:14 & SUB & 6 & ASIC \\
-- & -- & -- & -- & -- & 140905\_07:31:26 & EXT & none & ASIC \\
Per-emb-34 & 99 $\pm$ 13 & IRAS 03271+3013 & 03:30:15.12 & 30:23:49.20& 160917\_08:50:40\tablenotemark{g} & SUB & 2 & SWARM \\
-- & -- & -- & -- & -- & 160927\_08:02:56\tablenotemark{g} & SUB & 2, 3, 6 & SWARM \\
-- & -- & -- & -- & -- & 170122\_03:03:39 & SUB & 3 & SWARM \\
-- & -- & -- & -- & -- & 170122\_14:18:47 & SUB & 3 & SWARM \\
-- & -- & -- & -- & -- & 160208\_05:18:40 & COM & 4, 5, 7 & ASIC \\
-- & -- & -- & -- & -- & 160209\_03:40:33 & COM & 4, 5, 7 & ASIC \\
-- & -- & -- & -- & -- & 161030\_04:24:22 & EXT & 2 & SWARM \\
Per-emb-35 & 103 $\pm$ 26 & NGC 1333 IRAS~1 & 03:28:37.09 & 31:13:30.70 & 141213\_03:41:25 & SUB & 6 & ASIC \\
-- & -- & -- & -- & -- & 151006\_05:42:22 & EXT & 7 & both \\
Per-emb-36 & 106 $\pm$ 12 & NGC 1333 IRAS~2B & 03:28:57.36 & 31:14:15.70 & 151124\_03:10:17\tablenotemark{h} & SUB & none & ASIC \\
-- & -- & -- & -- & -- & 151129\_04:06:02\tablenotemark{h} & SUB & none & ASIC \\
-- & -- & -- & -- & -- & 161028\_05:14:31 & EXT & 2, 6 & SWARM \\
Per-emb-37 & 22 $\pm$ 1 & ... & 03:29:18.27 & 31:23:20.00 & 180911\_09:12:02 & SUB & 2 & SWARM  \\
% & 22 $\pm$ 1 & ... & 03:29:18.27 & 31:23:20.00 & 151203\_05:02:22 & SUB & none & ASIC  \\
-- & -- & -- & -- & -- & 161101\_04:33:39 & EXT & 2 & SWARM \\
Per-emb-38 & 115 $\pm$ 21 & ... & 03:32:29.18 & 31:02:40.90 & 170121\_04:28:59 & SUB & 3 & SWARM \\
-- & -- & -- & -- & -- & 161030\_04:24:22 & EXT & 2 & SWARM \\
Per-emb-39 & 125 $\pm$ 47 & ... & 03:33:13.78 & 31:20:05.20 & 160917\_08:50:40\tablenotemark{g} & SUB & 2 & SWARM \\
-- & -- & -- & -- & -- & 160927\_08:02:56\tablenotemark{g} & SUB & 2, 3, 6 & SWARM \\
-- & -- & -- & -- & -- & 170122\_03:03:39 & SUB & 3 & SWARM \\
-- & -- & -- & -- & -- & 170122\_14:18:47 & SUB & 3 & SWARM \\
-- & -- & -- & -- & -- & 160206\_03:00:59 & COM & 5, 7 & ASIC \\
-- & -- & -- & -- & -- & 161104\_04:21:48 & EXT & 2 & SWARM \\
Per-emb-40 & 132 $\pm$ 25 & B1-a & 03:33:16.66 & 31:07:55.20 & 151205\_04:33:28\tablenotemark{h} & SUB & none & ASIC \\
-- & -- & -- & -- & -- & 161028\_05:14:31 & EXT & 2, 6 & SWARM \\
Per-emb-41 & 157 $\pm$ 72 & B1-b & 03:33:20.96 & 31:07:23.80 & 141128\_03:49:43 & SUB & 6 & ASIC \\
-- & -- & -- & -- & -- & 150227\_03:14:20 & EXT & 5, 7 & ASIC \\
Per-emb-42 & 163 $\pm$ 51 & L1448C-S & \multicolumn{6}{c}{Imaged in the same pointing as Per-emb-26} \\
Per-emb-43 & 176 $\pm$ 42 & ... & 03:42:02.16 & 31:48:02.10 & 160925\_08:16:53\tablenotemark{g} & SUB & 2 & SWARM \\
-- & -- & -- & -- & -- & 160205\_03:21:56 & COM & 5, 7 & ASIC \\
-- & -- & -- & -- & -- & 161104\_04:21:48 & EXT & 2 & SWARM \\
Per-emb-44 & 188 $\pm$ 9 & SVS~13A & 03:29:03.42 & 31:15:57.72 & 151019\_06:11:24\tablenotemark{i} & SUB & 7 & ASIC \\
-- & -- & -- & -- & -- & 170127\_03:29:33 & SUB & 3 & SWARM \\
-- & -- & -- & -- & -- & 161031\_04:22:41 & EXT & 2 & SWARM \\
Per-emb-45 & 197 $\pm$ 93 & ... & 03:33:09.57 & 31:05:31.20 & 151205\_04:33:28\tablenotemark{h} & SUB & none & ASIC \\
-- & -- & -- & -- & -- & 161113\_03:41:32 & EXT & none & SWARM \\
Per-emb-46 & 221 $\pm$ 7 & ... & 03:28:00.40 & 30:08:01.30& 151108\_04:20:52 & SUB & none & ASIC \\
-- & -- & -- & -- & -- & 161028\_05:14:31 & EXT & 2, 6 & SWARM \\
Per-emb-47 & 230 $\pm$ 17 & IRAS 03254+3050 & 03:28:34.50 & 31:00:51.10 & 151019\_06:11:24\tablenotemark{i} & SUB & 7 &  ASIC \\
-- & -- & -- & -- & -- & 170127\_03:29:33 & SUB & 3 & SWARM \\
-- & -- & -- & -- & -- & 161108\_04:07:13 & EXT & none & SWARM \\
-- & -- & -- & -- & -- & 161110\_05:55:35 & EXT & none & SWARM \\
-- & -- & -- & -- & -- & 161112\_04:05:41 & EXT & none & SWARM \\
Per-emb-48 & 238 $\pm$ 14 & L1455-FIR2 & 03:27:38.23 & 30:13:58.80 & 151026\_05:33:00 & SUB & 7, 8 & ASIC \\
-- & -- & -- & -- & -- & 161102\_04:27:19 & EXT & 2 & SWARM \\
Per-emb-49 & 239 $\pm$ 68 & ... & 03:29:12.94 & 31:18:14.40& 141127\_02:21:26 & SUB & 6 & ASIC \\
-- & -- & -- & -- & -- & 150224\_04:47:03 & EXT & 5, 6, 7 & ASIC \\
-- & -- & -- & -- & -- & 150224\_05:33:03 & EXT & 5, 6, 7 & ASIC \\
-- & -- & -- & -- & -- & 151013\_04:46:09 & EXT & 7 & both \\
Per-emb-50 & 128 $\pm$ 23 & ... & 03:29:07.76 & 31:21:57.20 & 141127\_02:21:26  & SUB& 6 & ASIC \\
-- & -- & -- & -- & -- & 150224\_04:47:03 & EXT & 5, 6, 7 & ASIC \\
-- & -- & -- & -- & -- & 150224\_05:33:03 & EXT & 5, 6, 7 & ASIC \\
-- & -- & -- & -- & -- & 150915\_10:07:22 & EXT & 7 & ASIC \\
-- & -- & -- & -- & -- & 161029\_04:38:18 & EXT & 2 & SWARM \\
Per-emb-51 & 263 $\pm$ 115 & ... & 03:28:34.53 & 31:07:05.50 & 151026\_05:33:00 & SUB & 7, 8 & ASIC \\
-- & -- & -- & -- & -- & 161108\_04:07:13 & EXT & none & SWARM \\
-- & -- & -- & -- & -- & 161110\_05:55:35 & EXT & none & SWARM \\
-- & -- & -- & -- & -- & 161112\_04:05:41 & EXT & none & SWARM \\
Per-emb-52  & 278 $\pm$ 119 & ... & 03:28:39.72 & 31:17:31.90 & 151122\_11:23:42\tablenotemark{h} & SUB & none & ASIC \\
-- & -- & -- & -- & -- & 151122\_12:21:59\tablenotemark{h} & SUB & none & ASIC \\
-- & -- & -- & -- & -- & 151127\_04:06:10\tablenotemark{h} & SUB & none & ASIC \\
-- & -- & -- & -- & -- & 161102\_04:27:19 & EXT & 2 & SWARM \\
Per-emb-53 & 287 $\pm$ 8 & B5-IRS1 & 03:47:41.56 & 32:51:43.90 & 141130\_04:04:23 & SUB & 6 & ASIC \\
-- & -- & -- & -- & -- & 150204\_03:06:38 & EXT & 5, 6, 7 & ASIC \\
-- & -- & -- & -- & -- & 150912\_08:50:40 & EXT & 7 & ASIC \\
Per-emb-54 & 131 $\pm$ 63 & NGC 1333 IRAS~6 & 03:29:01.57 & 31:20:20.70 & 151022\_10:48:26 & SUB & 5, 7 & ASIC \\
-- & -- & -- & -- & -- & 180918\_08:32:35 & SUB & 6 & SWARM \\
-- & -- & -- & -- & -- & 161029\_04:38:18 & EXT & 2 & SWARM \\
Per-emb-55 & 309 $\pm$ 64 & IRAS 03415+3152 & \multicolumn{6}{c}{Imaged in the same pointing as Per-emb-8} \\
Per-emb-56 & 312 $\pm$ 1 & IRAS 03439+3233 & 03:47:05.42 & 32:43:08.40 & 141130\_04:04:23 & SUB & 6 & ASIC \\
-- & -- & -- & -- & -- & 151013\_04:46:09 & EXT & 7 & both \\
Per-emb-57 & 313 $\pm$ 200 & ... & 03:29:03.33 & 31:23:14.60 & 151206\_04:31:17\tablenotemark{h} & SUB & none & ASIC \\
-- & -- & -- & -- & -- & 161101\_04:33:39 & EXT & 2 & SWARM \\
Per-emb-58 & 322 $\pm$ 88 & ... & 03:28:58.44 & 31:22:17.40 & 151124\_03:10:17\tablenotemark{h} & SUB & none & ASIC \\
-- & -- & -- & -- & -- & 151129\_04:06:02\tablenotemark{h} & SUB & none & ASIC \\
-- & -- & -- & -- & -- & 161101\_04:33:39 & EXT & 2 & SWARM \\
Per-emb-59 & 341 $\pm$ 179 & ... & 03:28:35.04 & 30:20:09.90 & 151102\_04:48:11 & SUB & 7 & ASIC \\
-- & -- & -- & -- & -- & 161113\_03:41:32 & EXT & none & SWARM \\
Per-emb-60 & 363 $\pm$ 240 & ... & 03:29:20.07 & 31:24:07.50 & 151206\_04:31:17\tablenotemark{h} & SUB & none & ASIC \\
-- & -- & -- & -- & -- & 161102\_04:27:19 & EXT & 2 & SWARM \\
Per-emb-61 & 371 $\pm$ 107 & ... & 03:44:21.33 & 31:59:32.60 & 141130\_04:04:23 & SUB & 6 & ASIC \\
-- & -- & -- & -- & -- & 150204\_03:06:38 & EXT & 5, 6, 7 & ASIC \\
-- & -- & -- & -- & -- & 150912\_08:50:40 & EXT & 7 & ASIC \\
Per-emb-62 & 378 $\pm$ 29 & ... & 03:44:12.98 & 32:01:35.40 & 151123\_03:56:56\tablenotemark{h} & SUB & none & ASIC \\
-- & -- & -- & -- & -- & 151130\_04:08:59\tablenotemark{h} & SUB & none & ASIC \\
-- & -- & -- & -- & -- & 161029\_04:38:18 & EXT & 2 & SWARM \\
Per-emb-63 & 436 $\pm$ 9 & ... & 03:28:43.28 & 31:17:33.00 & 151122\_11:23:42\tablenotemark{h} & SUB & none & ASIC \\
-- & -- & -- & -- & -- & 151122\_12:21:59\tablenotemark{h} & SUB & none & ASIC \\
-- & -- & -- & -- & -- & 151127\_04:06:10\tablenotemark{h} & SUB & none & ASIC \\
-- & -- & -- & -- & -- & 161103\_06:11:02 & EXT & 2 & SWARM \\
Per-emb-64 & 438 $\pm$ 8 & ... & 03:33:12.85 & 31:21:24.10 & 151205\_04:33:28\tablenotemark{h} & SUB & none & ASIC \\
-- & -- & -- & -- & -- & 161103\_06:11:02 & EXT & 2 & SWARM \\
Per-emb-65 & 440 $\pm$ 191 & ... & 03:28:56.31 & 31:22:27.80 & 151124\_03:10:17\tablenotemark{h} & SUB & none & ASIC \\
-- & -- & -- & -- & -- & 151129\_04:06:02\tablenotemark{h} & SUB & none & ASIC \\
-- & -- & -- & -- & -- & 161103\_06:11:02 & EXT & 2 & SWARM \\
Per-emb-66 & 542 $\pm$ 110 & ... & 03:43:45.15 & 32:03:58.60 & 170121\_04:28:59 & SUB & 3 & SWARM  \\
-- & -- & -- & -- & -- & 160205\_03:21:56 & COM & 5, 7 & ASIC \\
-- & -- & -- & -- & -- & 161108\_04:07:13 & EXT & none & SWARM \\
-- & -- & -- & -- & -- & 161110\_05:55:35 & EXT & none & SWARM \\
-- & -- & -- & -- & -- & 161112\_04:05:41 & EXT & none & SWARM \\
B1-bN\tablenotemark{j} & 14.7 $\pm$ 1.0 & ... & 03:33:21.19 & 31:07:40.60 & 141128\_03:49:43 & SUB & 6 & ASIC \\
-- & -- & -- & -- & -- & 150227\_03:14:20 & EXT & 5, 7 & ASIC \\
B1-bS\tablenotemark{j} & 17.7 $\pm$ 1.0 & ... &  \multicolumn{6}{c}{Imaged in the same pointing as Per-emb-41} \\ 
L1448IRS2E\tablenotemark{j} & 15 & ... & 03:25:25.66 & 30:44:56.70 & 141129\_03:04:09 & SUB & 6 & ASIC \\
-- & -- & -- & -- & -- & 150923\_06:16:00 & EXT & 7, 8 & both \\
L1451-MMS\tablenotemark{j} & 15 & L1451-mm & 03:25:10.21 & 30:23:55.30 & 141129\_03:04:09 & SUB & 6 & ASIC \\
-- & -- & -- & -- & -- & 150923\_06:16:00 & EXT & 7, 8 & both \\
Per-bolo-45\tablenotemark{j} & 15 & ... & 03:29:07.70 & 31:17:16.80 & 141125\_04:39:14 & SUB & 6, 7, 8 & SWARM \\
-- & -- & -- & -- & -- & 170121\_04:28:59 & SUB & 3 & SWARM \\
-- & -- & -- & -- & -- & 140908\_06:06:05 & EXT & 6 & ASIC \\
-- & -- & -- & -- & -- & 150307\_02:24:55 & EXT & 5, 7 & ASIC \\
-- & -- & -- & -- & -- & 150307\_05:45:40 & EXT & 5, 7 & ASIC \\
Per-bolo-58\tablenotemark{j} & 15 & ... & 03:29:25.46 & 31:28:15.00 & 141125\_04:39:14 & SUB & 6, 7, 8 & ASIC \\
-- & -- & -- & -- & -- & 141214\_03:50:32 & SUB & 6 & ASIC \\
-- & -- & -- & -- & -- & 140908\_06:06:05 & EXT & 6 & ASIC \\
-- & -- & -- & -- & -- & 150307\_02:24:55 & EXT & 5, 7 & ASIC \\
-- & -- & -- & -- & -- & 150307\_05:45:40 & EXT & 5, 7 & ASIC \\
SVS~13B & 20 $\pm$ 20 & ... &  \multicolumn{6}{c}{Imaged in the same pointing as Per-emb-44} \\ 
SVS~13C & 21 $\pm$ 1 & ... & 03:29:01.97 & 31:15:38.05 & 151019\_06:11:24\tablenotemark{i} & SUB & 7 & SWARM \\
-- & -- & -- & -- & -- & 170127\_03:29:33 & SUB & 3 & SWARM \\
-- & -- & -- & -- & -- & 161031\_04:22:41 & EXT & 2 & SWARM 
\enddata
\tablenotetext{a}{The $T_{\text{bol}}$ values were taken from \citet{Tobin2016}. Sources with no errors were not detected by \emph{Herschel}, and \citet{Tobin2016} gave these sources approximate temperatures of 15~K.}
\tablenotetext{b}{Other names were taken directly from \citet{Tobin2016} and are not a complete list of other names for the target.}
\tablenotetext{c}{RA and DEC are given for the phase center of the observations. Accurate centers for envelopes are given in \citetalias{Stephens2018} and for protostars in \citet{Tobin2016}.}
\tablenotetext{d}{Only SUB and COM tracks had 850\,$\mu$m observations.}
\tablenotetext{e}{Each number represents the specific antenna number missing from the 8 antenna SMA array.}
%SUB tracks 160917\_08:50:40, 160925\_08:16:53, and 160927\_08:02:56 also do not have 850\,$\mu$m data. 151122\_11:23:42, 151123\_03:56:56, and 151124\_03:10:17}
\tablenotetext{f}{This track had a project code 2014B-S078, which is different than the rest of the tracks.}
\tablenotetext{g}{These SUB tracks do not have 850\,$\mu$m data.}
\tablenotetext{h}{These tracks are missing \cott\ and \hcop\ lines from the 850\,$\mu$m data.}
\tablenotetext{i}{This track was missing the ASIC chunks for \coto, \ttco, and the upper sideband s13.}
\tablenotetext{j}{This source is a first hydrostatic core candidate. More discussion on these sources are given in Section~\ref{sec:cfc}.}
%\tablenotetext{*}{These spectral lines that were not used in this study.} 
\end{deluxetable*}

The MASSES program (SMA project code 2014A-S093) observed the targets from 2014 to 2018, with the bulk of the observations done from 2014 to 2016. The name of the raw data file for each observing session is given in the sixth column of Table~\ref{tab:sources}. The format of the track name is YYMMDD\_HH:MM:SS (i.e., years, months, days, hours, minutes, seconds), which indicates the start time of the particular track. Some days have multiple tracks due to runtime issues, and these data were combined together during calibration. There were also additional tracks that were observed during the MASSES program that are not listed in Table~\ref{tab:sources}, but the data for those tracks were very poor and were not calibrated.\footnote{Track 151203\_05:02:22, which observed Per-emb-6, Per-emb-10, and Per-emb-37, was included in \citetalias{Stephens2018}, but it is excluded here as careful reanalysis of the data showed that the phases were too poor.} The SMA has eight antennas, and the eighth column of Table~\ref{tab:sources} indicates which, if any, antenna is missing from a particular track.

%SUB
%2014: 12 ASIC
%2015: 15 ASIC
%2016: 3 SWARM
%2017: 3 SWARM
%2018: 2 SWARM

%EXT
%2014: 5 ASIC
%2015: 9 ASIC, 4 both
%2016:  19 SWARM
During the time period in which MASSES was observed, the correlator was upgraded from ASIC (Application Specific Integrated Circuit) to SWARM  \citep[SMA Wideband Astronomical ROACH2 Machine;][]{Primiani2016}. During the commissioning of SWARM, both the ASIC and SWARM correlator could be used simultaneously with the SMA.
If we do not count same-day tracks as duplicates, in the SUB configuration a total of 27 tracks used ASIC and 8 used SWARM. For EXT, 14 used ASIC and 19 used SWARM, and 4 tracks used both ASIC and SWARM.\footnote{Several of the `ASIC-only' tracks actually used SWARM as well, but the SWARM data were corrupted and thus are not included in this tally.} The MASSES survey also observed four tracks using ASIC in SMA's compact configuration (COM). The configuration used for each track is given in Table~\ref{tab:sources}. The SUB configuration provides baselines from 9.5 to 77\,m, the COM configuration from 16 to 77\,m, and the EXT configuration from 44 to 226\,m. At 1.3\,mm, the observations are sensitive to scales of up to $\sim$30$\arcsec$ or 9000\,au. During imaging, we reach resolutions of up to $\sim$1$\arcsec$ or 300\,au (synthesized beams for each pointing are given in Table~\ref{tab:sens} and~\ref{tab:sens850}).

%In the SUB configuration, the shortest baseline was $\sim$9.5\,m and in the EXT configuration, the longest baseline was $\sim$226\,m. At 1.3\,mm, this allows for the observations to be sensitive to scales of $\sim$1.2$\arcsec$ to 29$\arcsec$, or 360\,au to 8700\,au at the distance to Perseus.

%230 GHz:
%USB: ~230.1
%LSB: ~220.7
%(220.7+230.1)/2 = 225.4
%c/225.4 GHz = 1.33 mm

%345 GHz: 
%USB is centered at about 355.9
%LSB is centered at about 346.6
%(355.9+346.6)/2 = 351.25
%c/351.25 GHz = 853.5 um

\renewcommand{\tabcolsep}{0.1cm}
\begin{deluxetable*}{lcccccccc}
\tablecolumns{3}
\tabletypesize{\scriptsize}
\tablewidth{0pt}
\tablecaption{Primary Spectral Lines Covered by the MASSES Survey \label{tab:lines}}
\tablehead{\colhead{Tracer} & \colhead{Transition} & \colhead{Frequency} & \colhead{ASIC} & \colhead{ASIC Channels} & \colhead{$\Delta v_{uv,\rm{ASIC}}$\tablenotemark{a}} & \colhead{$\Delta v_{uv,\rm{SWARM}}$\tablenotemark{a}} & \colhead{$\Delta v_{img}$\tablenotemark{a}} & \vspace{-8pt} \\ %\colhead{Number of Imaged} \\
& & \colhead{(GHz)} & \colhead{Chunk} & \colhead{Per Chunk} & \colhead{(\kms)} & \colhead{(\kms)} & \colhead{(\kms)} &  %\colhead{Channels}
}
\startdata
1.3\,mm cont & & 231.29\tablenotemark{b} & LSB s05 -- s12, s14 & 64\tablenotemark{c} & & & \\ %& 1 \\
& & & USB s05 -- s12 \\
\cotont & \emph{J} = 2 -- 1 & 230.53796 & USB s13, s14\tablenotemark{d} & 512 & 0.26 & 0.18 & 0.5 \\% & ~\,220/430\tablenotemark{e}\\
\ttcont & \emph{J} = 2 -- 1 & 220.39868 & LSB s13 & 512 & 0.28 & 0.19 & 0.3 \\ %& 200\\
\ceont & \emph{J} = 2 -- 1 & 219.56036 & LSB s23 & 1024 & 0.14 & 0.19 & 0.2 \\ %& 200\\
\ntdpnt & \emph{J} = 3 -- 2 & 231.32183 & USB s23 & 1024 & 0.13 & 0.18 & 0.2 \\ %& ~125
%\smallskip \\ 
\hline  
850\,$\mu$m cont & & 356.72/356.410\tablenotemark{e}  & LSB, USB s04 -- s12 & 64\tablenotemark{c} \\
& & & USB s05 -- s08, s10 -- s12 \\
\cottnt & \emph{J} = 3 -- 2 & 345.79599 & LSB s18 & 512 & 0.088 & 0.12 & 0.5 \\
\hcopnt & \emph{J} = 4 -- 3 & 356.73424 & USB s18 & 1024 & 0.085 & 0.12 & 0.2 \\
\httcopnt & \emph{J} = 4 -- 3 & 346.99835 & LSB s04 & 1024 & 0.088 & 0.12 & 0.2
\enddata
%\tablecomments{Rows in italics are spectral lines that were not used in this study}
\tablenotetext{a}{Velocity resolutions $\Delta v_{uv}$ and $\Delta v_{img}$ are for the $uv$ data and imaged data, respectively.}
\tablenotetext{b}{Tuning frequencies for the 1.3\,mm SMA observations. Based on the chunks used for the continuum, the continuum frequency for the ASIC correlator is closer to 225\,GHz. One track, 160927\_08:02:56, had a different tuning frequency of 230.538\,GHz. ASIC and SWARM tracks have a total 1.3\,mm continuum bandwidth of 1.394\,GHz and up to $\sim$16\,GHz, respectively. }
\tablenotetext{c}{The channel width for the LSB s14 and s04 chunks is 512 and 1024~channels, respectively. }
\tablenotetext{d}{The central velocity and the majority of the \coto\ line is in the s14 chunk. The s13 chunk contains higher, positive velocities.}
\tablenotetext{e}{Tuning frequencies for the SMA observations.  The first value is for ASIC, and the second is for SWARM. Based on the chunks used for continuum, the continuum frequency for the ASIC correlator is closer to 351\,GHz. ASIC and SWARM tracks have a total 1.3\,mm continuum bandwidth of 1.312\,GHz and up to $\sim$16\,GHz, respectively.}
\end{deluxetable*}

The SMA can simultaneously observe at a different frequency for each of its two receivers. For the SUB and COM tracks, we tuned the receivers to observe at 1.3\,mm and 850\,$\mu$m simultaneously, while for EXT tracks, we only tuned receivers to 1.3\,mm. The tuning frequencies for the receivers with ASIC data were 231.29 and 356.72\,GHz and for SWARM data were 231.29 and 356.41\,GHz. The ASIC correlator provides a bandwidth of 2\,GHz in each sideband, and each sideband is divided into 24~$\times$~104\,MHz chunks. The frequencies of the chunks slightly overlap, so that the ``effective" bandwidth of a chunk is 82\,MHz. The spectral resolution depends on how many channels are assigned to each chunk, which is described in more detail in \citetalias{Stephens2018}.  The spectral lines specifically targeted are given in Table~\ref{tab:lines}. This table also provides information on which chunk(s) was (were) used for the continuum and each spectral line (with the indication of the lower and/or upper sideband, i.e., LSB or USB), as well as the velocity resolution for each spectral line. At 1.3\,mm and 850\,$\mu$m, the continuum bandwidth was 1.394 and 1.312\,GHz, respectively.

 \begin{deluxetable}{lcccccccc}
\tablecolumns{3}
\tabletypesize{\scriptsize}
\tablewidth{0pt}
\tablecaption{Other SWARM Lines Detected toward Some Pointings\label{tab:swarmlines}}
\tablehead{\colhead{Tracer} & \colhead{Transition} & \colhead{Frequency} & \colhead{$E_u$} \\
& & \colhead{(GHz)} & \colhead{(K)}
%\tablehead{\colhead{Tracer} & \colhead{Transition} & \colhead{Frequency} & \colhead{ASIC} & \colhead{ASIC Channels} & \colhead{$\Delta v_{uv,\rm{ASIC}}$\tablenotemark{a}} & \colhead{$\Delta v_{uv,\rm{SWARM}}$\tablenotemark{a}} & \colhead{$\Delta v_{img}$\tablenotemark{a}} & \colhead{Number of Imaged} \\
%& & \colhead{(GHz)} & \colhead{Chunk} & \colhead{Per Chunk} & \colhead{(\kms)} & \colhead{(\kms)} & \colhead{(\kms)} & \colhead{Channels}
}
\startdata
SO & $J_N = 5_5 -4_4$ & 215.22065 & 44.1 \\
DCO$^+$ & $J = 3 - 2$ & 216.11258 & 20.7 \\
SiO & $J=5 - 4$ & 217.10498 & 31.3 \\
DCN	& $J = 3 -2$ & 217.23854 & 20.9 \\
c-C$_3$H$_2$ & $6_{0,6}$--5$_{1,5}$ & 217.82215 & 38.6 \\
c-C$_3$H$_2$ & $6_{1,6}$--5$_{0,5}$ & 217.82215 & 38.6 \\
H$_2$CO & $J_{K_a,K_b} = 3_{0,3} - 2_{0,2}$ & 218.22219 & 21.0\\
H$_2$CO & $J_{K_a,K_b} = 3_{2,2} - 2_{2,1}$ & 218.47563 & 68.1\\
H$_2$CO & $J_{K_a,K_b} = 3_{2,1} - 2_{2,0}$ & 218.76007 & 68.1\\
SO & $J_N = 6_5 -5_4$ & 219.94944 & 35.0
\smallskip \\ 
\hline  
$^{34}$SO & $J_N = 9_8- 8_7$ &  339.85763 & 77.3 \\
%SO & $J_N = 3_3 - 3_2$ & 339.34145900 & 9.23 \\
OCS & $J = 28-27$ & 340.44927 & 220.6 \\
HC$^{18}$O$^+$ & $J = 4-3$ & 340.63069 & 40.9 \\
SO & $J_N = 7_8 - 6_7$ & 340.71416 & 64.9 \\
CS & $J = 7-6$ & 342.88285 & 65.8 \\
HC$^{15}$N & $J = 4-3$ & 344.20032 & 41.3 \\
SO & $J_N = 8_8 - 7_7$ & 344.31061 & 71.0 \\
%SO & $J_N = 2_3 - 2_1$ & 345.70455530 & 4.46 \\
H$^{13}$CN & $J = 4-3$ & 345.33977 & 41.4 \\
SO & $J_N = 9_8 - 8_7$ & 346.52848 & 62.1 \\
CN & $J=5/2-3/2$ & & 32.6 \\
& $F=5/2-5/2$ & 340.00813 \\
& $F=3/2-3/2$ & 340.01963 \\
& $F=7/2-5/2$ & 340.03155\tablenotemark{a} \\
& $F=3/2-1/2$ & 340.03541 \\
& $F=5/2-3/2$ & 340.03541 \\
CN & $J = 7/2-5/2$ & & 32.7 \\
& $F=7/2-5/2$ & 340.24777\tablenotemark{a} & \\
& $F=9/2-7/2$ & 340.24777 & \\
& $F=5/2-3/2$ & 340.24854 & \\
& $F=5/2-5/2$ & 340.26177 & \\
& $F=7/2-7/2$ & 340.26495 & \\
DCO$^+$ & $J=5-4$ & 360.16978 & 51.9 \\
HNC & $J=4-3$ & 362.63030 & 43.5\ \\
DCN & $J=5-4$ & 362.04648 & 52.1 
\enddata
\tablecomments{Splatalogue (\url{http://www.splatalogue.net/}) is used for the frequencies and upper energy levels. Other lines may also exist within the delivered visibility data. Images of these spectral lines are only provided for targets with at least one SWARM SUB track.}
\tablenotetext{a}{The delivered cube covers all of these $F$ transitions for the given $J$ transition. The delivered CN cube has this rest frequency in the fits header.}
\end{deluxetable}

The SWARM correlator provides up to 16\,GHz bandwidth per sideband, which is divided into four equal-sized chunks. The entire bandwidth has a uniform spectral resolution of 140\,kHz (0.18\,\kms\ at 233\,GHz). Because observations were taken while the correlator was being commissioned, sometimes we were unable to use the full (four~chunk) SWARM correlator, so the bandwidth was less than 16\,GHz; for example, most EXT SWARM observations had only a 12\,GHz bandwidth per sideband. The bandwidth available with SWARM is significantly higher than it is for ASIC, providing for a significant improvement for the continuum sensitivities. Because the SWARM correlator has a large bandwidth with uniformly high spectral resolution, we serendipitously detected many additional spectral lines when this correlator was used. These lines are only easily identified using the SUB configuration, but as mentioned above, only eight SUB tracks (targeting 18 unique protostars) used the SWARM correlator. We have manually identified the additional spectral lines that are detected toward at least some of these protostars, which are listed in Table~\ref{tab:swarmlines}. All of these lines listed in this table were detected in the SVS~13 system (i.e., MASSES protostars Per-emb-44, SVS~13B, and SVS~13C). While considerable effort was made to manually identify these spectral lines, more lines could be detected toward at least some protostars, especially if one uses more advanced spectral line identification techniques.

As mentioned above, 850\,$\mu$m observations were only observed in the SUB and COM configurations. In some situations, the 850\,$\mu$m data were not delivered for a particular track because either there were correlator problems or the weather was too poor to allow for sufficient calibration (the atmospheric opacity tends to be three times higher at 850\,$\mu$m than 1.3\,mm for the same precipitable water vapor). In other situations, \cott\ and \hcop\ ASIC chunks were dropped from the correlator due to a power supply failure. Nevertheless, each MASSES target has successful 850\,$\mu$m continuum and \httcop\ observations. Two targets, Per-emb-7 and Per-emb-43, had 850\,$\mu$m data that were taken in the COM configuration only and were missing antennas. Per-emb-7 850\,$\mu$m image data products have particularly poor image fidelity and point source and temperature sensitivities, as only five antennas were in the array and integration times were short.

\begin{figure}[ht!]
\begin{center}
\includegraphics[width=1\columnwidth]{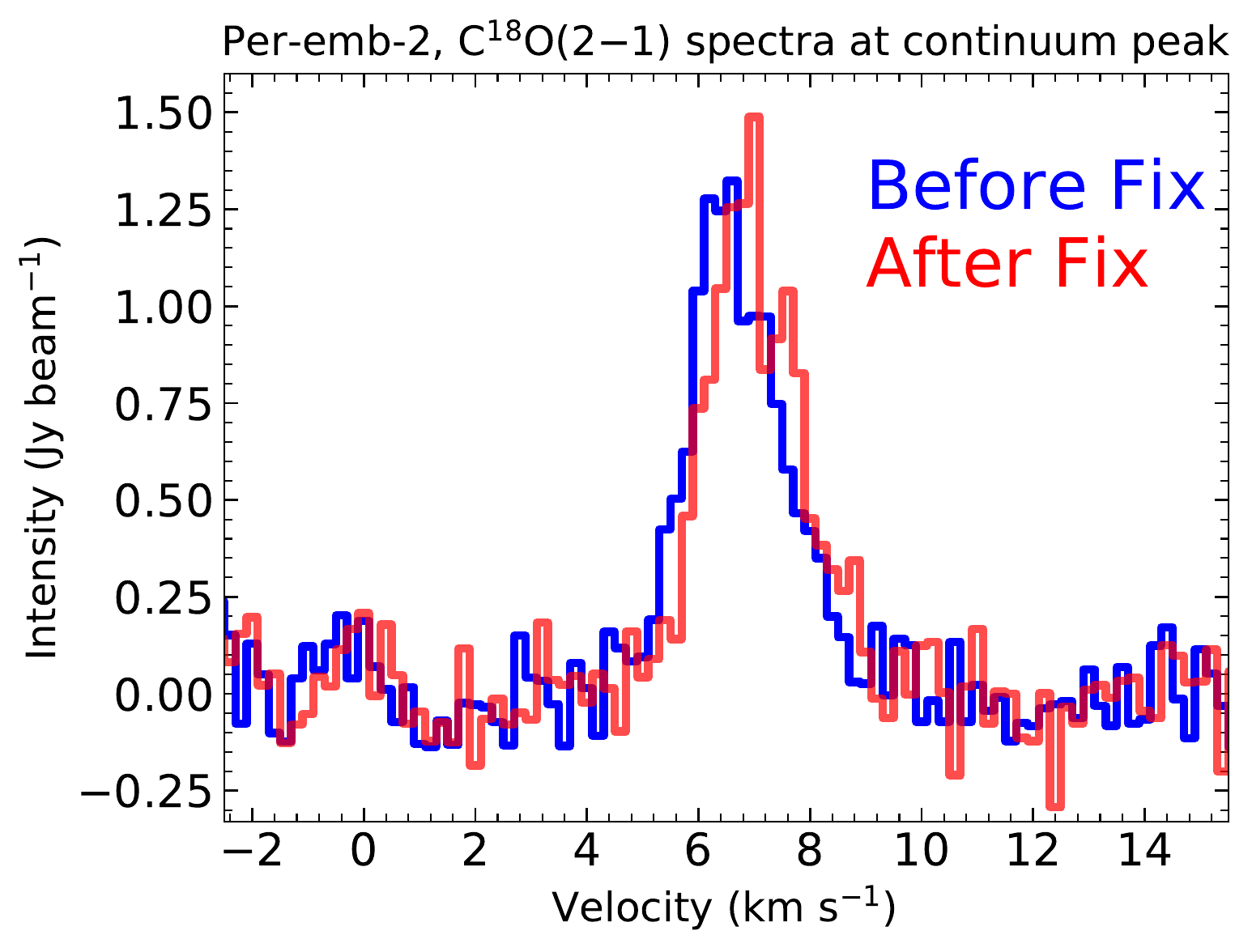}
\includegraphics[width=1\columnwidth]{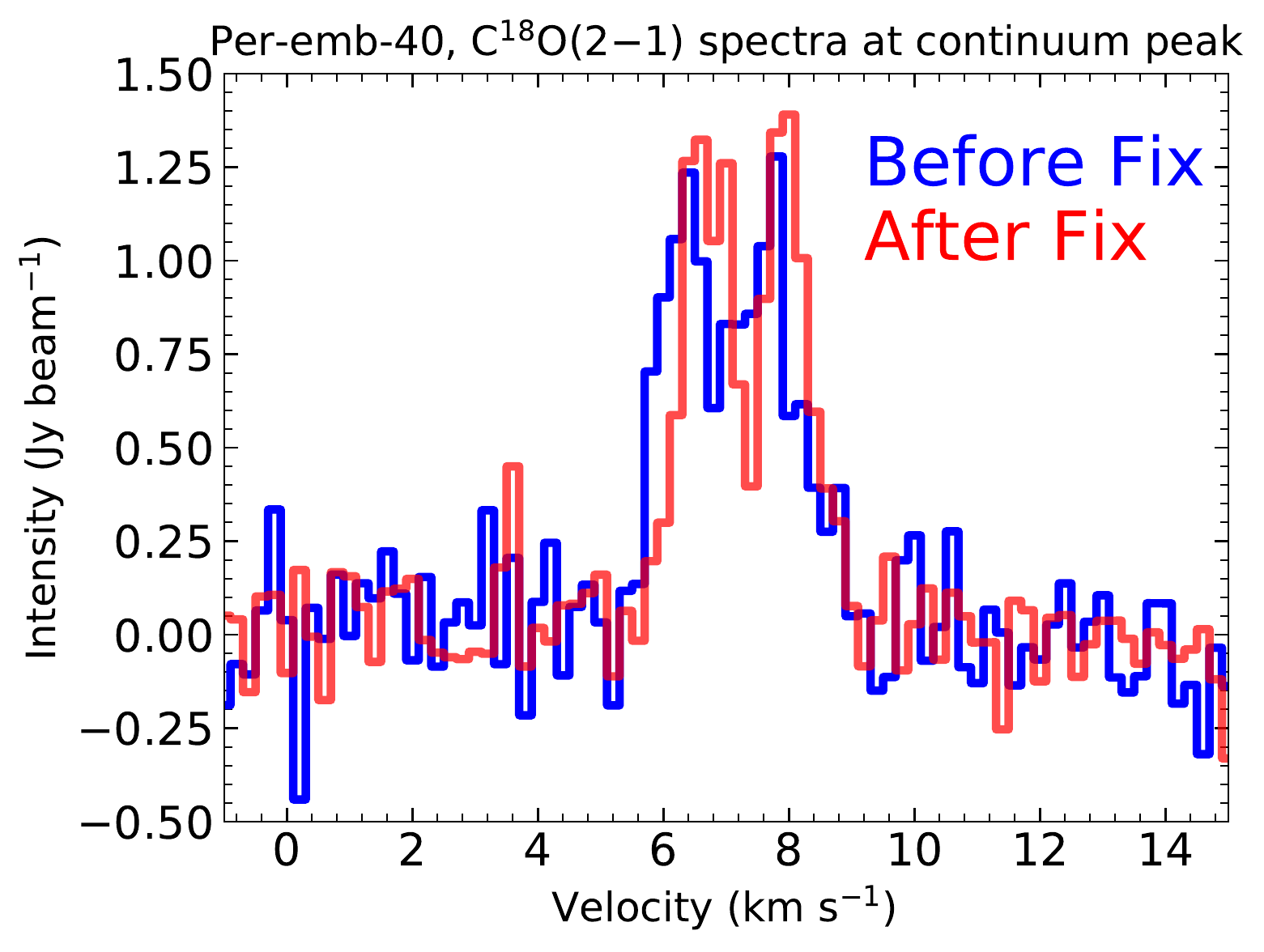}
\end{center}
\caption{Primary-beam-corrected \ceo\ spectra, taken at the 1.3\,mm continuum peak pixel for Per-emb-2 (top panel) and Per-emb-40 (bottom panel). The blue spectra show what the spectra looked before the fix, while the red spectra show after the fix.
}
\label{fig:hayshft} 
\end{figure}

\section{Data Processing}
\subsection{Data Calibration}\label{sec:datacal}
All data calibration was performed with the MIR software package, which is described in the MIR cookbook.\footnote{\url{https://www.cfa.harvard.edu/~cqi/mircook.html}} The data reduction process for the EXT 230 GHz and 345 GHz data is identical to the data reduction process of the SUB 230 GHz data, described fully in \citetalias{Stephens2018}. The one exception is that the data were further corrected for a recently discovered SMA real-time system software error controlling Doppler tracking as a function of time. The Doppler tracking error, described in detail in a forthcoming SMA Memo,\footnote{\url{https://www.cfa.harvard.edu/sma/memos/}} affected all data from mid-2011 until April 2019. In essence, the diurnal portion of the line-of-sight velocity (due to Earth's rotation) was calculated for the wrong location.  Therefore, each scan (30\,s for MASSES) was assigned an incorrect diurnal motion Doppler correction, with the error changing in magnitude through the track. The net effect when using the full track's unfixed data is that (1) spectral features are smoothed up to $\sim$1\,\kms\ in velocity space, and (2) the peak spectra has a small shift of order a few tenths of a \kms. In other words, the error introduces both smearing and a velocity bias/offset.  The SMA staff developed a new MIR task called \texttt{uti\_hayshft\_fix}, which is run on calibrated data to remove the errant Doppler correction as well as input the correct shift, on a scan by scan basis. To correct the shift for all sources in each track, the task \texttt{uti\_doppler\_fix} is then called. Applying this calibration technique is straightforward and is described in the MIR cookbook. We applied this new correction to every calibrated dataset for the MASSES program, including the previously published data described in \citetalias{Stephens2018}.

The effect of the correction is significant for Perseus protostars given the relatively narrow lines (on order of 1\,\kms). We found that over the course of a track for Perseus, the Doppler velocity correction for each integration could amount to differences of up to $\sim$1\,\kms.  In comparison of pre- and post-correction data at locations where the spectral lines are bright, flux differences in each channel of up to 50\% were not uncommon. The correction, by removing the effects of the smearing, significantly improved the sharpness of all narrow spectral line features, allowing for easier identification of weak narrows lines and higher signal-to-noise ratio for stronger ones. In Figure~\ref{fig:hayshft}, the products before and after applying the Doppler correction fix are shown for \ceo\ spectra at a single pixel for Per-emb-2 and Per-emb-40. It is apparent that without the fix, the velocity has a significant shift, and the spectral features are smoothed. Because the real-time error was not identified until after the publication of \citetalias{Stephens2018}, the results presented there were adversely affected; we stress that the data presented in this paper improves on and supersedes that presented in \citetalias{Stephens2018}. After the Doppler shift correction was applied, all data were converted to MIRIAD format \citep{Sault1995}.

 \subsection{Data Imaging}\label{sec:imaging}
All imaging was done using MIRIAD and is also described in detail in \citetalias{Stephens2018}. We repeat most of the details here, as there are a couple of minor differences which will be specified below. 

For spectral lines, the continuum is first subtracted using task \texttt{uvlin}, where a range of line-free channels (visually identified) is specified by the user. The task fits a DC offset to these channels and subtracts the offset from the window. For both continuum and spectral lines, an inverse Fourier transform was used on all visibilities (i.e., specifying all tracks for each target as specified in Table~\ref{tab:sources}) to create a dirty image using the MIRIAD task \texttt{invert} and specifying options ``systemp" and ``double". During the invert task, we create two data products of different resolutions by specifying Brigg's robust parameters of both --1 and 1. This differs from the subcompact release \citepalias{Stephens2018}, where we only delivered products that use robust~=~1.  A robust parameter of 2 is similar to natural weighting, which minimizes the noise in the image, while a robust parameter of --2 is comparable to uniform weighting, which yields the highest resolution. Each weighting has an advantage, and an intermediate robust parameter allows for a balance between the two. We find that robust parameters of --1 and 1 are both good balances, so each is delivered.

We then clean the dirty images via the MIRIAD task \texttt{clean}. We use a three-step cleaning process for \coto, \ttco, \cott, and \hcop, which are lines that typically trace the protostellar outflows. This is described in more detail in \citetalias{Stephens2018}, but we briefly summarize it here. During the first step, we manually select pixels and channels with known emission and clean these to a level of 1.5~times the dirty map noise. The manual selection was performed in the robust~=~1 maps by binning channels with similar emission structure, creating moment 0 maps of each bin, and selecting the pixels for these maps using the task \texttt{cgcurs}. For the second step, we cleaned over all user-specified pixels that had at least some emission throughout the entire position--position--velocity cube to a level of 2~times the dirty map noise. To select these pixels, we use the  robust~=~1 maps to make a single moment 0 map over the entire range in which there is line emission and again select pixels using \texttt{cgcurs}. The same pixels were used for cleaning both robust-weighting data products. For the third step, all pixels and channels were cleaned to a level of 2.5~times the dirty map noise. For the continuum and all other spectral lines (including SWARM-only lines), we only implement a two-step iterative cleaning algorithm, eliminating the second step described above. Moreover, for any continuum or spectral line data where we found that no emission was obviously associated with sources in the pointing (i.e., the tracer was undetected or there were interferometric artifacts from large-scale emission), we did not apply any iterative cleaning algorithm, but rather simply cleaned all pixels and channels in the dirty map to 3~times the dirty map noise. The clean components were then used with the dirty map and beam to create cleaned maps via the MIRIAD task \texttt{restor}. Data were not self-calibrated, as we found that self -alibration only had very marginal, if any, improvement on the data (see \citetalias{Stephens2018} for details).

In the subcompact data release \citepalias{Stephens2018}, we specified ``options=positive" when running the \texttt{clean} task, which constrains the clean component image to be nonnegative. This option was initially used as it appeared to bring out low-level emission. However, after testing this option, we found that it frequently adds a significant positive bias to any pixel that is cleaned, causing erroneously higher fluxes. In this full release, we do not use this option.

For both robust weightings, the sensitivities and the dimensions of the synthesized beams for each target are given in tables in the appendix. Table~\ref{tab:sens} shows these for 1.3\,mm, and Table~\ref{tab:sens850} shows these for 850\,$\mu$m observations. To keep the appendix concise, we do not include the sensitivities for the SWARM-only lines. Sensitivities and beam sizes can vary throughout each of SWARM's 16\,GHz sideband, but in general, the sensitivity and beams in the imaged 0.2\,\kms\ SWARM channels are similar to the main spectral lines imaged at the same spectral resolutions, i.e., \ceo, \ntdp, \hcop, and \httcop. The sensitivities can be easily calculated by the user by opening the data cubes and measuring the sensitivities in the line-free channels. When mapping the 1.3\,mm continuum for 23 of the 68 pointings, specifying a robust parameter of --1 did not create higher resolution images than when using a robust parameter of 1. All 23 of these pointings had SUB tracks that used the ASIC correlator only and used at least one EXT track with SWARM. Therefore, the longer baselines already had higher sensitivity due to the extra continuum bandwidth provided by SWARM. For these 23 pointings, the robust~=~--1 maps are not useful as their sensitivity are worse than the robust~=~1 maps and are not at higher resolution. As such, we leave these entries empty in Table~\ref{tab:sens}, and we do not release these targets 1.3\,mm continuum products for the robust~=~--1 weighting.

The angular extents of the delivered cubes/images are 80$\arcsec$ $\times$ 80$\arcsec$, which is considerably bigger than the full width at half maximum (FWHM) of the primary beam (48$\farcs$0 at 230\,GHz and 31$\farcs$2 at 350\,GHz). We map over this large angular extent as there is bright emission frequently detected well outside the FWHM of the primary beam.

Table~\ref{tab:detection} lists whether the continuum and the main spectral lines (i.e., those listed in Table~\ref{tab:lines}) are detected toward each pointing. The designation of whether or not a line is detected was determined ``by eye". A spectral line was considered detected if there was an obvious spectral peak within the cube, or if there was a weak peak but it was located near the protostar at the systemic velocity. The detection was considered marginal if either the peak was extremely weak but at the correct position and velocity, or if the peak was slightly weak but the line was not at the exact expected position/velocity. Note that Table~\ref{tab:detection} only specifies whether these tracers are detected within the pointing, regardless of whether it is associated with a protostar.  While every CO isotopolog is detected toward each pointing, sometimes the emission is not associated with the protostar at all. These lines are always detected because they are bright lines permeating the large scales of Perseus, and the interferometer cannot completely filter them out.

\subsection{Continuum Mass Sensitivity}\label{sec:mass_sensitivity}
We follow \citetalias{Stephens2018} to determine the mass sensitivity of the observations, as described below. For optically thin dust emission, the flux of a source can be translated into a gas mass via

\begin{equation} \label{m_disk_eq}
	M = R_{\text{gd}} \frac{F_{\nu} d^2}{\kappa_{\nu} B_{\nu}(T_{\text{dust}})},
\end{equation}
where $R_{\text{gd}}$ is the gas to dust mass ratio (assumed to be 100), $F_\nu$ is the flux of the source, $d$ is the distance to the source (300\,pc), $\kappa_{\nu}$ is the dust opacity, and $B_{\nu}(T_{\text{dust}}$) is the Planck function for dust temperature $T_{\text{dust}}$. We use the value $\kappa_{\rm{1.3\,mm}}$~=~0.899\,cm$^2$\,g$^{-1}$, following \citet{Ossenkopf1994} with the assumption of thin ice mantles and a gas density of 10$^6$\,cm$^{-3}$ (i.e., the so-called OH5 opacities). To be conservative with the minimum detectable mass, we assume $T_{\text{dust}}$~=~10\,K and a 3$\sigma$ detection of $F_{\nu}$~=~3($\sigma_{\rm{1.3\,mm}} \times \text{bm})$. The latter assumption for $F_{\nu}$ assumes that the source is a point source at the center of the primary beam. If the source is extended or significantly offset from the center, the mass sensitivity limit is underestimated. We arrive at

\begin{equation} \label{m_disk_eq_limit}
	M_{\text{limit, 1.3\,mm}} = \left(\frac{\sigma_{\rm{1.3\,mm}}}{\rm{mJy\,bm}^{-1}}\right) \times 0.016\,M_\sun ,
\end{equation}
where $\sigma_{\rm{1.3\,mm}}$ is the sensitivity given in Table~\ref{tab:sens}. Note that this value is higher than that reported in \citetalias{Stephens2018} because Perseus has been updated to a farther distance of 300~pc. Similarly, we can calculate the mass sensitivity for the 850\,$\mu$m observations. Following the same assumptions above with $\kappa_{850\,\mu \rm{m}}$~=~1.84\,cm$^2$\,g$^{-1}$ \citep{Ossenkopf1994}, we arrive at

%2.57*(700/852)^1.7=1.84
\begin{equation} \label{m_disk_eq_limit850}
	M_{\text{limit, 850\,$\mu$m}} = \left(\frac{\sigma_{850\,\mu \rm{m}}}{\rm{mJy\,bm}^{-1}}\right) \times 0.0048\,M_\sun ,
\end{equation}
where $\sigma_{850\,\mu \rm{m}}$ is the sensitivity given in Table~\ref{tab:sens850}.

The continuum sensitivity across the entire MASSES sample varies dramatically (by a factor of $\sim$80) due to dynamic range limitations, whether or not the protostar was observed with the SWARM correlator, and the quality and amount of tracks observed toward a target. Per-emb-4 is an example of a source where we were not dynamic range limited (no continuum source detected), and we reached sensitivities of 0.0075\,$M_\sun$ and 0.032\,$M_\sun$ for the 1.3\,mm and 850\,$\mu$m observations, respectively. Given that most 1.3\,mm observations have additional EXT SWARM tracks and the 850\,$\mu$m observations are usually ASIC SUB tracks only, the mass sensitivity is typically much better for the 1.3\,mm continuum.

\section{Data Products}
%The description of the data products is quite similar to \citetalias{Stephens2018}. We summarize the most important details about the data products below.

All MASSES data are publicly available on dataverse via \url{https://dataverse.harvard.edu/dataverse/full_MASSES/}. The MASSES dataverse contains two datasets for each of the 68 pointings. One dataset contains the $uv$ (visibility) data, while the other contains the imaged data. A README is also available in each dataset that briefly describes ways to use the dataset.

\subsection{uv Data}
The $uv$~data are delivered for each source, with one dataset per track for each spectral line and two datasets per track for the continuum (one each for the LSB and USB). Datasets from Table~\ref{tab:sources} with the same YYMMDD prefix were combined in a single track in MIR, so they are delivered as a single track. The delivered $uv$ data for spectral lines have been continuum subtracted. 

We also deliver $uv$ data for the entire SWARM bandwidth for all eight (four for each sideband) spectral chunks. These data were not continuum subtracted, as an accurate continuum subtraction is done better near each spectral line rather than across an entire 4~GHz chunk. This is due to many factors, including the fact that the atmospheric transmission can vary significantly over the entire SWARM bandwidth, and protostellar envelopes have a significant spectral index. A user splicing spectra from the $uv$ data will need to do their own continuum subtraction, if needed.

For both ASIC and SWARM chunks, the edge channels (especially the 1200 edge channels of each SWARM chunk) can have considerable noise. These channels should either be used with extreme caution or thrown out entirely (as they are with our imaged data).

All data are delivered in uvfits format. The format of the spectral line $uv$~data is\\

\indent ~~~NAME.LINE.TRACK.uvfits,\\

\noindent where NAME is an abbreviated name. For example, Per-emb-5 becomes Per5, and Per-emb-18 and Per-emb-21, which were mapped in the same pointing, becomes Per18Per21. LINE corresponds to the spectral line observed, with its transition usually indicated (e.g., 12CO21, 13CO21, etc.). The TRACK is in the track number prefix, which is in the format of YYMMDD.

The format of the continuum $uv$~data is\\

\indent ~~~NAME.cont$\lambda$.SB.TRACK.uvfits,\\

\noindent where $\lambda$ is either `1.3mm' or `850um' and the SB is either `lsb' or `usb.'  If SWARM $uv$ data are available, the format is\\

\indent ~~~NAME.SWARM.$\lambda$.CHUNK.SB.TRACK.uvfits,\\

\noindent where CHUNK is s1, s2, s3, or s4. Examples of these three formats are shown for the Per-emb-3 track 180918\_08:32:35:

\begin{enumerate}
\item Per3.12CO21.180918.uvfits
\item Per3.cont1.3mm.lsb.180918.uvfits
\item Per3.SWARM.850um.s3.usb.180918.uvfits.
\end{enumerate}

\subsection{Imaged Data}
We deliver continuum-subtracted cubes for each spectral line observation and a single 2D image for each continuum observation. We deliver both images/cubes that have and have not been corrected for the primary beam response. Images not corrected for the primary beam are useful for displaying structure across the entire field. Cube/images corrected for the primary beam response should be used if the user wants to extract accurate fluxes. The format for the images/cubes delivered without and with primary correction is\\

\indent  NAME.LINE.robustNUM.fits,\\
\indent  NAME.LINE.robustNUM.pbcor.fits, \\

\noindent respectively, where LINE in this instance can also include continuum, and NUM is either --1 or 1, depending on the robust parameter used for the imaging. Examples for Per-emb-3 images are:

\begin{enumerate}
\item Per3.12CO21.robust1.fits
\item Per3.C18O21.robust1.pbcor.fits
\item Per3.cont850um.robust-1.pbcor.fits.
\end{enumerate}

The continuum images and spectral line cubes are in units of Jy\,bm$^{-1}$ and Jy\,bm$^{-1}$\,channel$^{-1}$, respectively. Spectral line images are delivered as cubes with the velocity resolution specified in Table~\ref{tab:lines}. SWARM lines are all imaged with a spectral resolution of 0.2\,\kms, except for SiO(5--4), which is also imaged at 0.5\,\kms. We provide images of the additional SWARM lines listed in Table~\ref{tab:swarmlines} only if the target has at least one SWARM SUB track, as the sensitivity is too poor for these lines for EXT-only tracks. Nevertheless, $uv$ data are provided in case the user desires to look at additional EXT SWARM lines in more detail.

%Total three sigmas: 10
%Median linewidth of 3-sigmas: 0.5160866767621866
%[0.4943831596918237, 0.3307666852206077, 0.4695945986658476, 0.5280596878856413, 0.3489970895594818, 0.7042230745658827, 0.504113665638732, 0.8363667829272653, 0.6192838501431865, 0.8975660171493277]
%Total two sigmas: 23
%Median linewidth of 2-sigmas: 0.5744596096872179
%[0.4943831596918237, 0.7448911961429625, 0.44186546235930624, 1.6492010948596925, 0.3307666852206077, 0.4695945986658476, 0.3847181624815869, 0.2949608532197345, 0.8309375971034548, 0.5280596878856413, 0.3378394749668691, 0.7068356467734953, 0.6046539972093135, 0.3489970895594818, 0.7042230745658827, 0.504113665638732, 0.6950167810268508, 0.9733529272055926, 0.8363667829272653, 0.5744596096872179, 0.6192838501431865, 0.3113939789560055, 0.8975660171493277]

\section{Survey Overview}
\subsection{Detection Statistics}
As mentioned in Section~\ref{sec:imaging}, we show in Table~\ref{tab:detection} which spectral lines were detected toward each pointing, and the CO isotopologues are detected toward every pointing, even if the emission is not necessarily associated with the protostar. \hcop\ was almost always detected, i.e., toward 47 of the 53 (89\%) pointings. In \citetalias{Stephens2018}, we showed that \ntdp\ is significantly more likely to be detected toward the youngest protostars. This trend is also found for \httcop. As can be seen by examining Table~\ref{tab:detection}, any time \httcop\ is detected, \ntdp\ is also detected. \ntdp\ is definitively detected toward 42 of the 68 pointings (62\%), with an additional 2 marginal detections. \httcop\ was detected toward 32 of the 68 pointings (47\%), with an additional 5 marginal detections. At 1.3\,mm and 850\,$\mu$m, the continuum was detected in 58 (85\%) and 51 (75\%) of the 68 pointings, respectively, with 3 and 5 marginal detections, respectively. For the continuum and the main spectral lines (i.e., those listed in Table~\ref{tab:lines}), there is emission detected for 487 of the 582 (84\%) of the delivered robust~=~1 images/cubes.
%3 CO lines: 3*68 of 3*68, or 204 of 204
%CO(3-2): 53 of 53
%HCO+: 47 of 53
%N2D+: 42 of 68
%H13CO+: 32 of 68
%1.3 cont: 58 of 68
%870 um cont: 51 of 68
%204+53+47+42+32+58+51 = 487
%204+53+53+68+68+68+68 = 582
%6+25+36+10+17= 94

\begin{figure*}[ht!]
\begin{center}
\includegraphics[width=1.9\columnwidth]{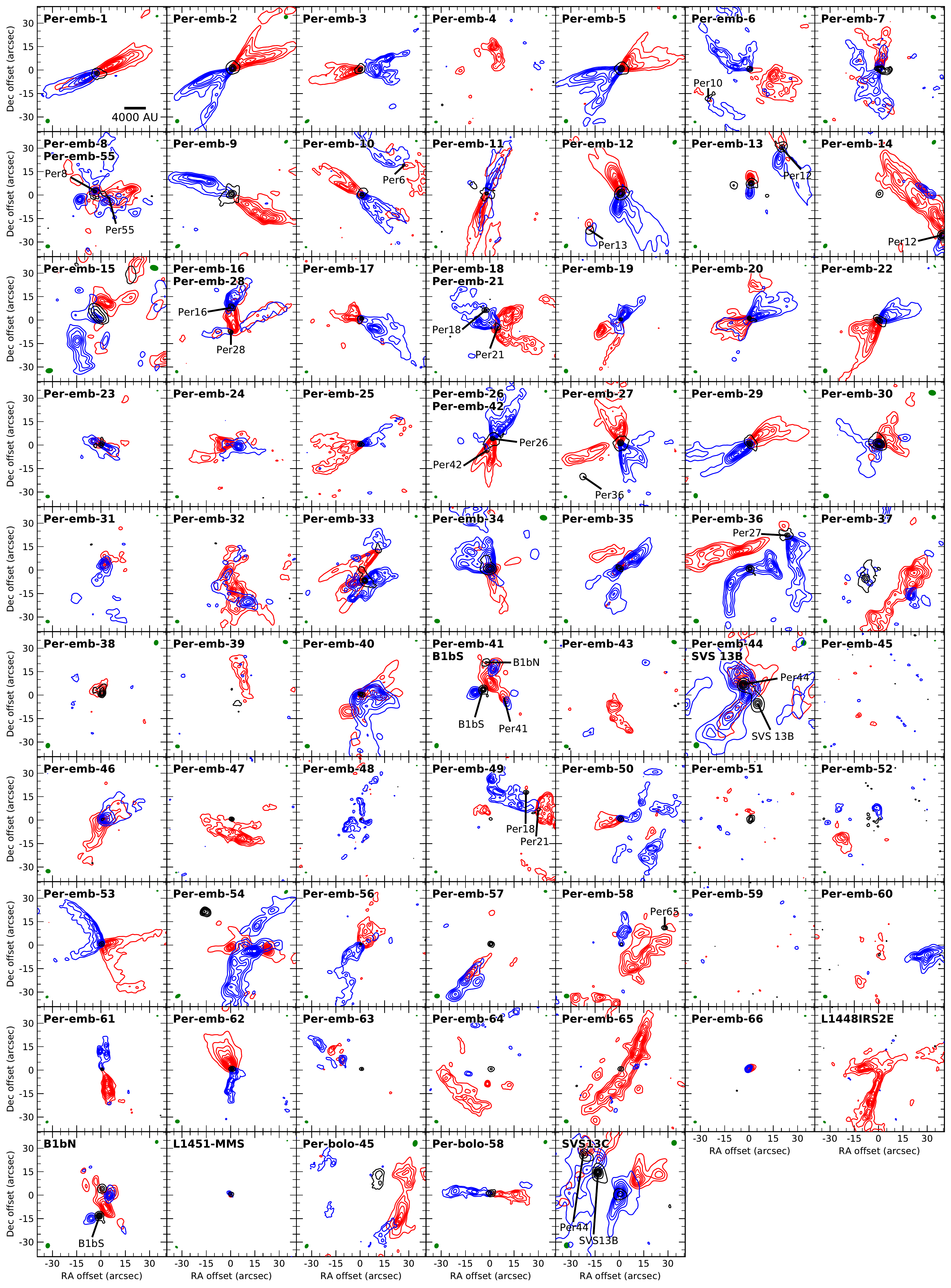}
\end{center}
\vspace{-18pt}
\caption{\coto\ integrated intensity (moment 0) and continuum maps for all pointings, imaged using the robust~=~1 weighting. Black contours show the 1.3\,mm continuum, while blue and red show the integrated intensity (moment~0) maps for the blue- and red-shifted emission, respectively. The contours and velocity integration ranges are given in Table~\ref{tab:contours}. The top-right and bottom-left green ellipses show the synthesized beams for the continuum and \coto, respectively. Fluxes have not been corrected for the primary beam.
}
\label{fig:rp1outflows} 
\end{figure*}

\begin{figure*}[ht!]
\begin{center}
\includegraphics[width=1.9\columnwidth]{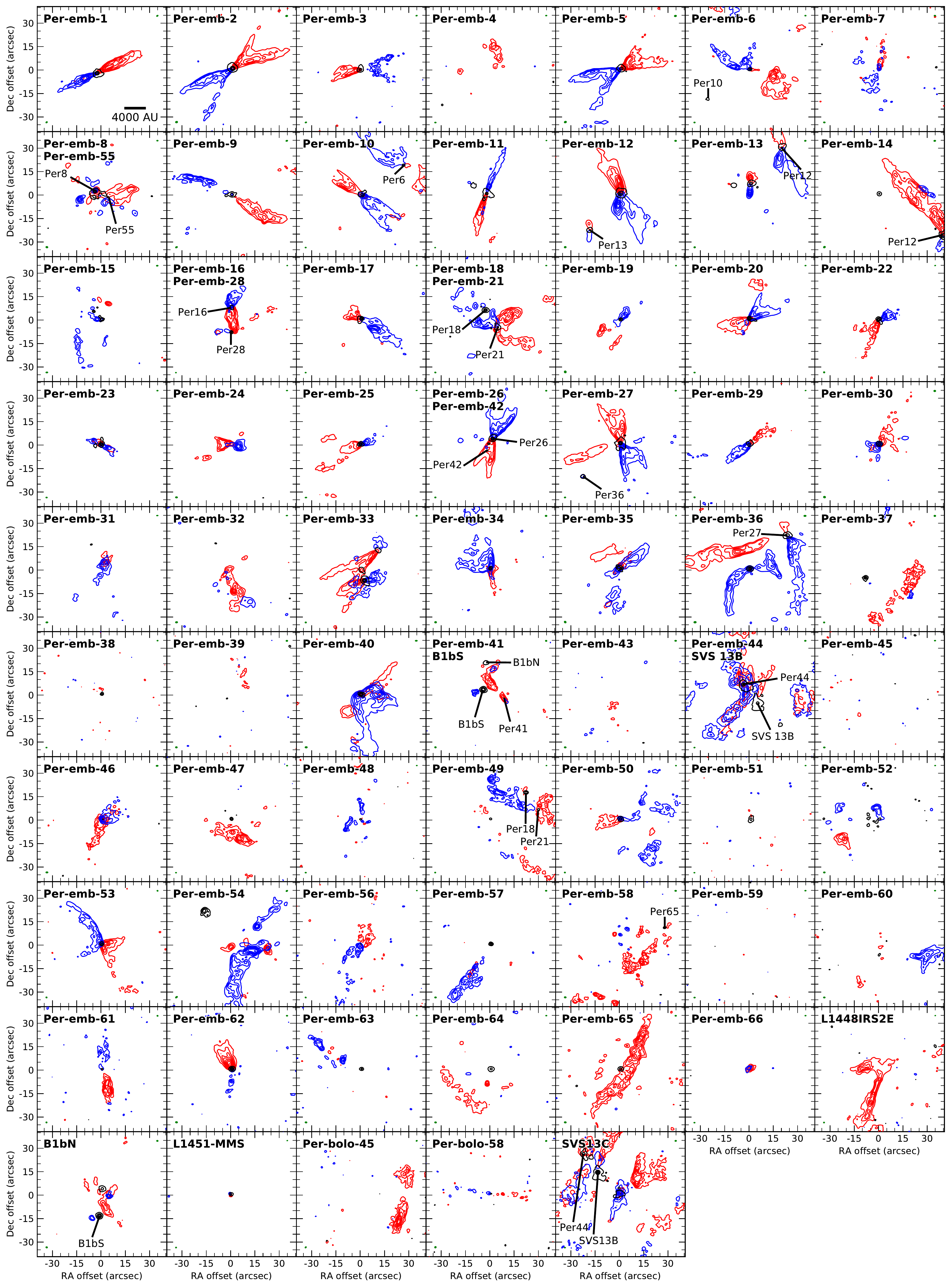}
\end{center}
\caption{Same as Figure~\ref{fig:rp1outflows}, but now using robust~=~--1 (i.e., lower sensitivity but higher resolution). The 1.3\,mm continuum maps without a corresponding robust~=~--1 image (see Section~\ref{sec:imaging} and Table~\ref{tab:sens}) use the robust~=~1 image.
}
\label{fig:rp-1outflows} 
\end{figure*}

%\begin{figure*}[ht!]
%\begin{center}
%\includegraphics[width=2\columnwidth]{robust1_MASSES_continuum.pdf}
%\end{center}
%\caption{Continuum maps for all pointings, imaged using robust~=~1 weighting. Green contours are the same as the black contours shown in Figure~\ref{fig:rp1outflows}, and the contours are given in Table~\ref{tab:contours}. The top right blue ellipse shows the synthesized beam. Yellow stars indicate protostars detected by the VLA via the VANDAM survey \citep{Tobin2016}. Fluxes have not been corrected for the primary beam.
%}
%\label{fig:rp1cont} 
%\end{figure*}

\begin{figure*}[ht!]
\begin{center}
\includegraphics[width=1.9\columnwidth]{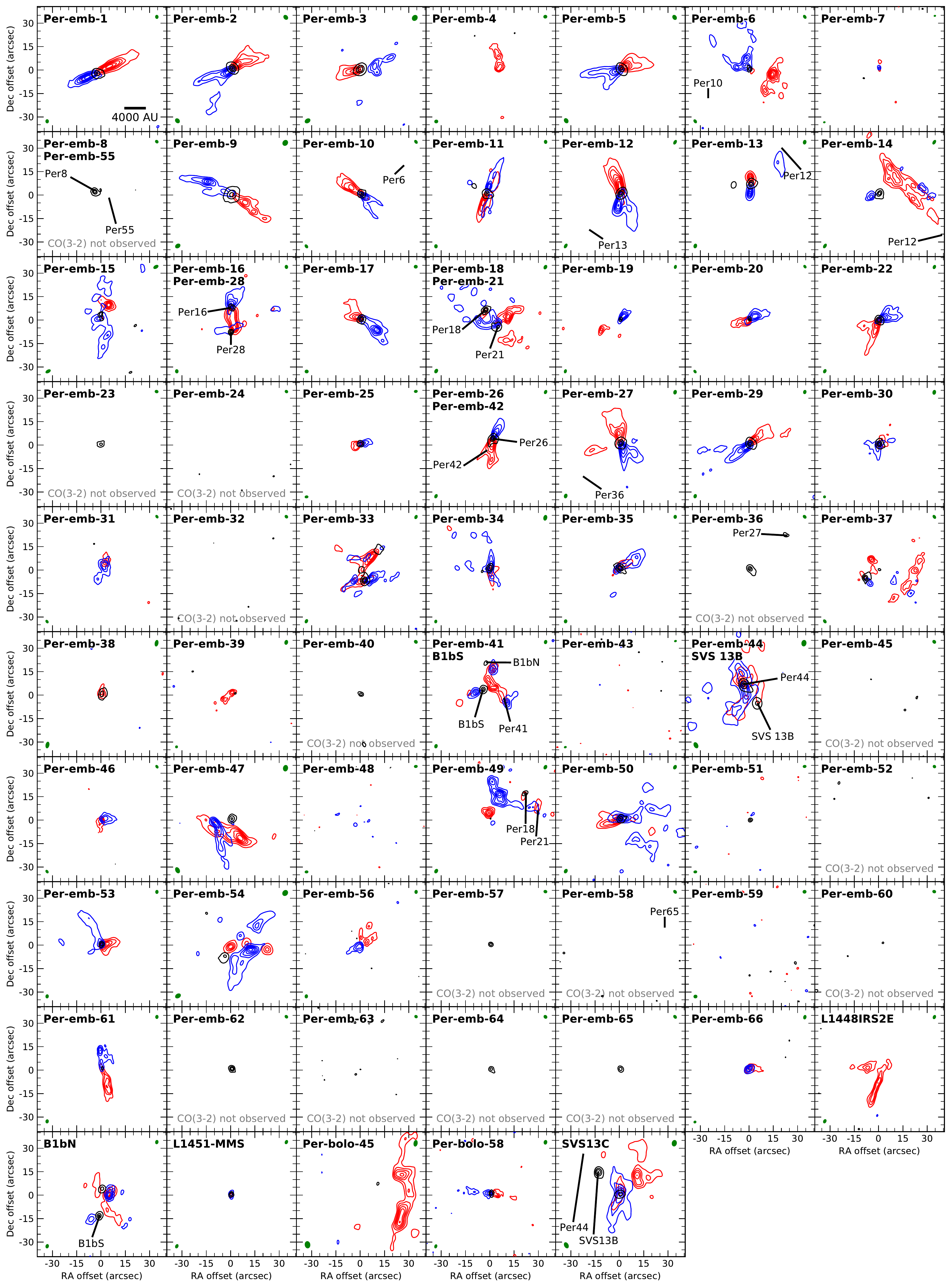}
\end{center}
\caption{Same as Figure~\ref{fig:rp1outflows}, but now using the 850\,$\mu$m continuum and \cott\ maps. Contour levels are given in Table~6. Note that \cott\ was not observed toward some pointings.
}
\label{fig:co32outflows} 
\end{figure*}

\vspace{-6pt}
\begin{figure*}[ht!]
\begin{center}
\includegraphics[width=1.9\columnwidth]{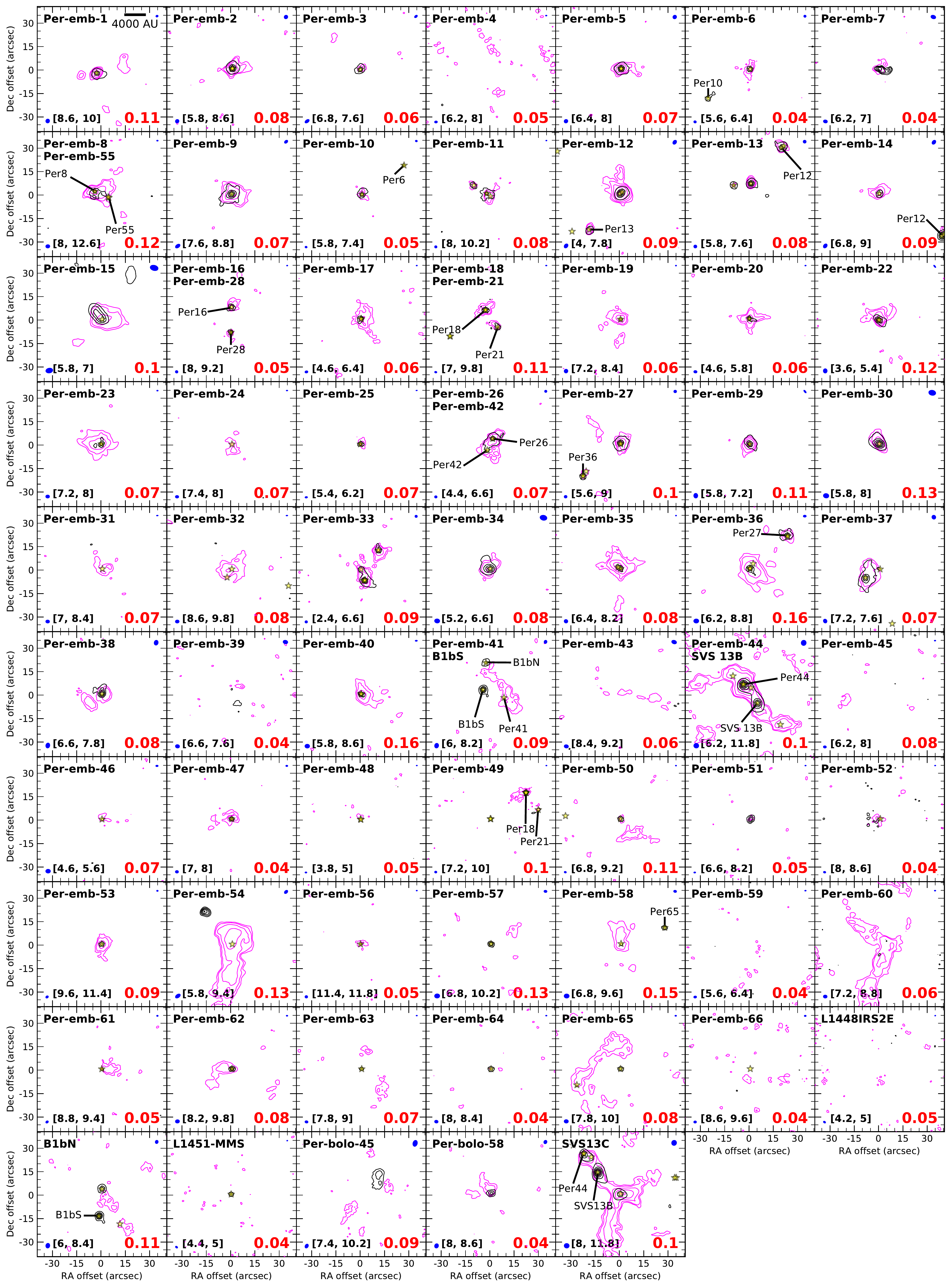}
\end{center}
\vspace{-18pt}
\caption{Continuum maps (black contours) overlaid with \ceo\ (magenta contours) for all pointings, imaged using robust~=~1 weighting. Black continuum contours are the same as those in Figure~\ref{fig:rp1outflows}, with the contour levels given in Table~\ref{tab:contours}. The magenta \ceo\ contours have levels of $f$~$\times$~[3,5,10,20,40,60], where the value for $f$ is indicated in red in the bottom right of each panel. The integrated velocity ranges are indicated in brackets at the bottom left of each panel where [start,~stop] indicate the start and stop of the velocity integration range in units of \kms. The top-right and bottom-left green ellipses show the synthesized beams for the continuum and \ceo, respectively. Yellow stars indicate protostars detected by the VLA via the VANDAM survey \citep{Tobin2016}. Fluxes have not been corrected for the primary beam.
}
\label{fig:contceo} 
\end{figure*}

\vspace{-6pt}
\begin{figure*}[ht!]
\begin{center}
\includegraphics[width=1.9\columnwidth]{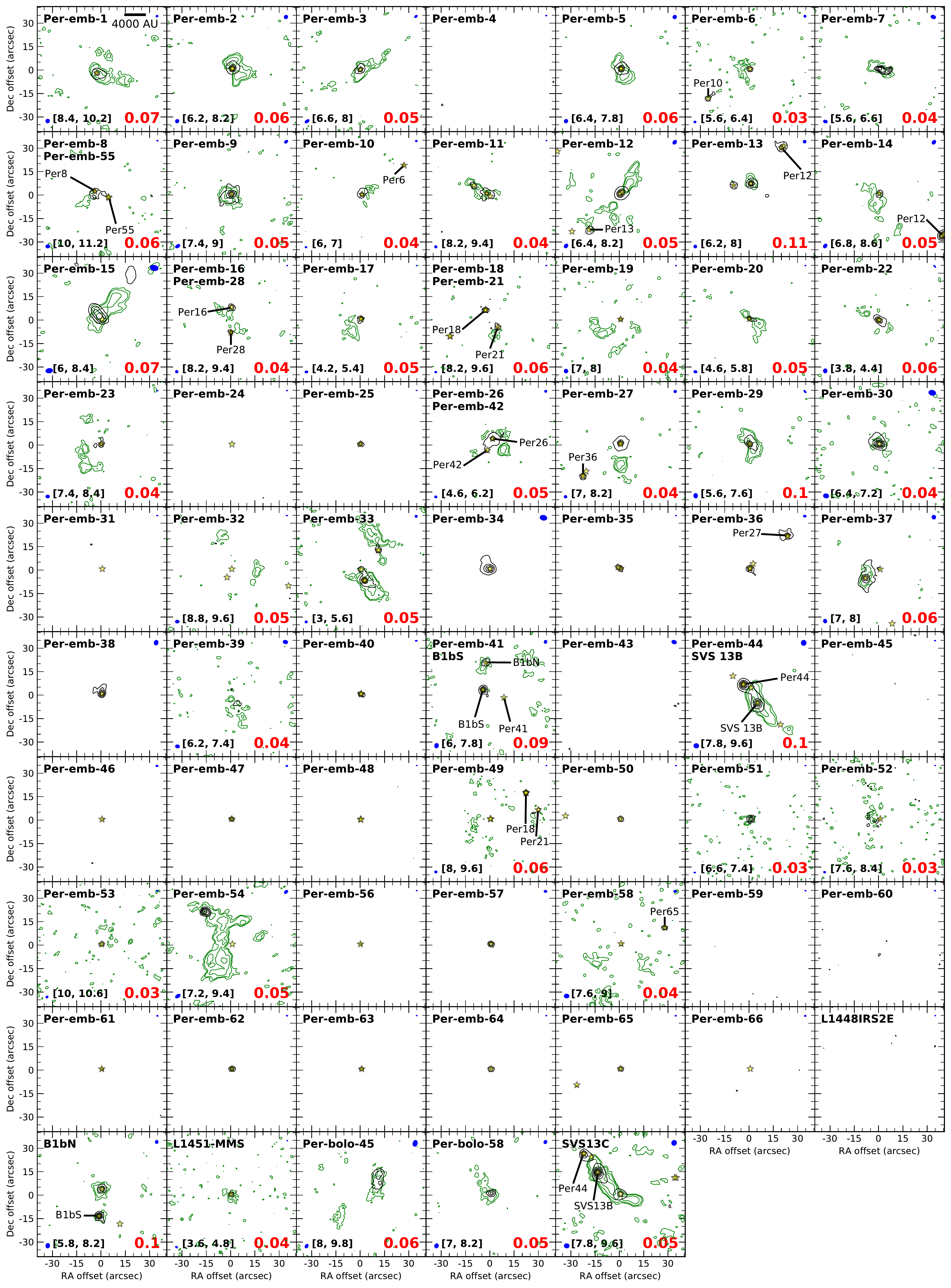}
\end{center}
\vspace{-18pt}
\caption{Continuum maps (black contours) overlaid with \ntdp\ (green contours, shown for detections only) for all pointings, imaged using robust~=~1 weighting. Black continuum contours are the same as those in Figure~\ref{fig:rp1outflows}, with the contour levels given in Table~\ref{tab:contours}. The green \ntdp\  contours have levels of $f$~$\times$~[3,5,10,20], where the value for $f$ is indicated in red in the bottom right of each panel. The integrated velocity ranges are indicated in brackets at the bottom left of each panel where [start,~stop] indicate the start and stop of the velocity integration range in units of \kms. The top-right and bottom-left green ellipses show the synthesized beams for the continuum and \ntdp, respectively. Yellow stars indicate protostars detected by the VLA via the VANDAM survey \citep{Tobin2016}. Fluxes have not been corrected for the primary beam.
}
\label{fig:contn2dp} 
\end{figure*}

%\subsection{Continuum and CO Outflows}
\subsection{Spectral Line and Continuum Images}
Here we present the images of the primary spectral lines of the MASSES survey (i.e., those from Table~\ref{tab:lines}). We present multiple galleries of protostellar CO emission, which primarily trace bipolar outflows, along with the continuum. Figure~\ref{fig:rp1outflows} and~\ref{fig:rp-1outflows} show imaged \coto\ and 1.3\,mm continuum for the 68 pointings using robust weightings robust~=~1 and robust~=~--1, respectively. Both weightings have their advantages, as the robust~=~1 images are more sensitive and capture the more extended structure, while the robust~=~--1 maps resolve finer structures. Compared to what was mapped with the SUB-only configuration in \citetalias{Stephens2018}, we noticed that the outflows have considerable substructure, with many local peaks. Such local peaks are also noticed in the 0.5\,\kms\ channels. We believe these substructures are real and are not due to imaging artifacts, as we found that these substructures are apparent in the dirty maps. Figure~\ref{fig:co32outflows} shows the imaged \cott\ with a weighting of robust~=~1. Only 53 pointings had \cott, as many SUB tracks were missing \cott\ due to correlator problems (as discussed in Section~\ref{sec:setup}; see Table~\ref{tab:sources}). Generally, bipolar outflows for both \coto\ and \cott\ are clearer for protostars in the upper panels, which are typically the younger protostellar sources (i.e., those with lower bolometric temperatures). These protostars are likely more entrained in gas and have more active accretion that drives these outflows.

Figures~\ref{fig:contceo} and~\ref{fig:contn2dp} show \ceo\ and \ntdp, respectively, mapped with the 1.3\,mm continuum. These spectral lines primarily trace the protostellar envelopes. Protostars detected by the VANDAM survey are also overlaid in these figures.

%c18o_spectra_full.py
\begin{figure*}[ht!]
\begin{center}
\includegraphics[width=2\columnwidth]{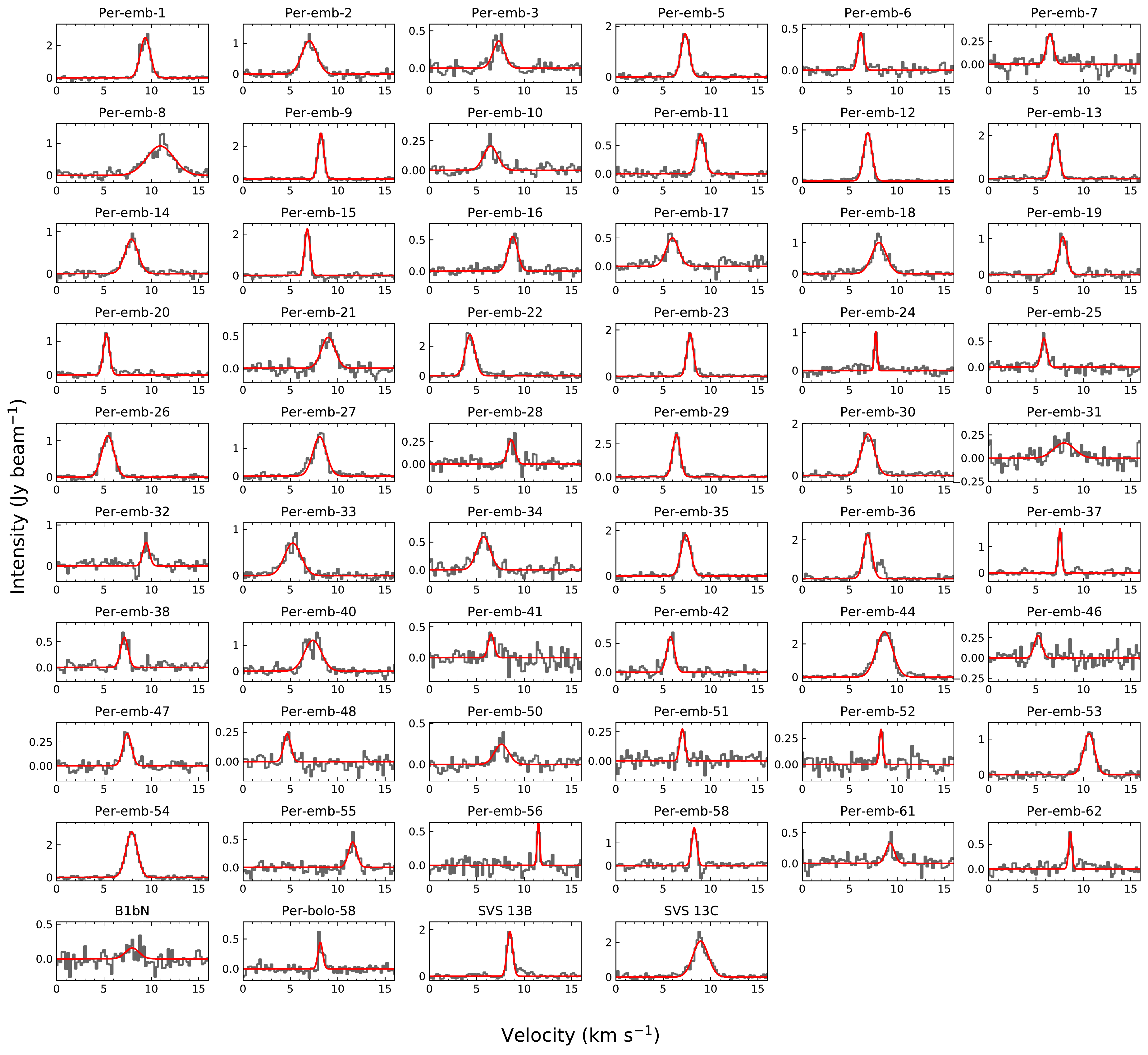}
\end{center}
\caption{Fitted primary-beam-corrected \ceo\ spectra of MASSES protostars of the robust~=~1 cubes. Spectra are toward the peak of each protostellar envelope and averaged for an aperture with a radius of 1$\farcs$2 (3\,pixels). A single-component Gaussian fit is shown as a red curve, with the parameter of the fit given in Table~\ref{tab:fits}.
}
\label{fig:ceospectra} 
\end{figure*}

\subsection{\ceo\ Spectra}
We fit the primary beam corrected, robust~=~1 \ceo\ spectra toward each protostar with a single-component Gaussian. In general, \ceo\ is better than \ntdp\ for measuring linewidths toward the protostar itself as \ntdp\ can often be anticorrelated with the continuum emission, which is readily seen by comparing Figures~\ref{fig:contceo} and \ref{fig:contn2dp}, which is probably due to increased temperature toward the protostar \citep[e.g.,][]{Emprechtinger2009,Tobin2013c}. We fit spectra only when the \ceo\ emission visually appears to be associated with the protostar, which is based on visual analysis of \ceo\ cubes and their integrated intensity maps. The spectra is extracted at the location of the peak of the 1.3\,mm continuum (location from MIRIAD's \texttt{maxfit} task), taking the average spectra using an aperture with a radius of 3~pixels = 1$\farcs$2. This aperture size was chosen primarily because it generally had a negligible change from the single-pixel fit for most protostars while allowing for the \ceo\ spectra of other protostars to be more apparent (most notably Per-emb-48). The spectra and their fits are shown in Figure~\ref{fig:ceospectra}, and the Gaussian fit parameters (amplitude, systemic velocity $v_{\text{systemic}}$, and the FWHM linewidth $\Delta v$) are given in Table~\ref{tab:fits}. Protostars that do not obviously have \ceo\ associated with them are not included in the figure or table. 

While we only fit the spectra with a single component Gaussian, there are some spectra that notably have two peaks (e.g., Per-emb-36 and Per-emb-40). Without careful analysis (and the need for more data toward some protostars), we cannot judge whether these two peaks are due to multiple components, self-absorption in the spectra, or confusion (e.g., due to large-scale emission or outflows). As such, we only use the single-component Gaussian fits. 

The median linewidth measured for a MASSES protostar is 1.0\,\kms. As mentioned in Section~\ref{sec:datacal}, one substantial improvement of these data over the subcompact data release is the doppler correction. This doppler error had the net effect of smoothing the data over velocity. The subcompact data release found 1.45\,\kms\ for the median linewidth, which is higher than the 1.0\,\kms\ values measured here. While some of this difference is due to this the doppler tracking error, other factors (e.g., different size scales probed and more signal in each spectra with the large beam in \citetalias{Stephens2018}) contribute to this difference as well. As mentioned in \citetalias{Stephens2018}, the \ceont\ linewidths measured with single-dish telescopes at both 1$\arcmin$ (0.09\,pc) and 11$\arcsec$ (3300\,au) resolution are typically $\sim$0.6--1.0\,\kms\ \citep{Hatchell2005,Kirk2007}. The median 1.0\,\kms\ measured in the MASSES sample are at the high end of the single-dish range, suggesting the typical linewidths are slightly larger. The larger linewidths we measure at the envelope scale may be due to the increased kinematic activity apparent at these scales (e.g., outflows, rotation, and infall).

%c18o_spectra_full.py
\startlongtable
\begin{deluxetable}{lcccc}
%\tablecolumns{20}
%\tabletypesize{\scriptsize}
%\tablewidth{-20pt}
\tablecaption{ \ceo\ Fitting Information \label{tab:fits}}
\tablehead{ %&  \multicolumn{3}{c}{\underline{~~~~~~\ceo\ Spectra Fit\tablenotemark{c}~~~~~}} \vspace{-5pt} \\
\colhead{Source\tablenotemark{a}}  & Amplitude & $v_{\text{systemic}}$ & \colhead{$\Delta v$\tablenotemark{a}}  \vspace{-8pt} \\
\colhead{Name} & (Jy\,bm$^{-1}$) & (\kms) & (\kms) 
} 
\startdata
Per-emb-1 & 2.5 $\pm$ 0.05 & 9.4 $\pm$ 0.01 & 1.2 $\pm$ 0.03 \\
Per-emb-2 & 1.1 $\pm$ 0.04 & 7.0 $\pm$ 0.03 & 1.8 $\pm$ 0.07 \\
Per-emb-3 & 0.36 $\pm$ 0.03 & 7.3 $\pm$ 0.06 & 1.4 $\pm$ 0.1 \\
Per-emb-5 & 1.7 $\pm$ 0.04 & 7.3 $\pm$ 0.01 & 1.0 $\pm$ 0.03 \\
Per-emb-6 & 0.46 $\pm$ 0.03 & 6.2 $\pm$ 0.02 & 0.69 $\pm$ 0.05 \\
Per-emb-7 & 0.34 $\pm$ 0.04 & 6.5 $\pm$ 0.05 & 0.87 $\pm$ 0.11 \\
Per-emb-8 & 0.92 $\pm$ 0.03 & 11.0 $\pm$ 0.06 & 3.3 $\pm$ 0.1 \\
Per-emb-9 & 2.8 $\pm$ 0.04 & 8.2 $\pm$ 0.006 & 0.74 $\pm$ 0.01 \\
Per-emb-10 & 0.21 $\pm$ 0.02 & 6.4 $\pm$ 0.07 & 1.7 $\pm$ 0.2 \\
Per-emb-11 & 0.70 $\pm$ 0.02 & 9.0 $\pm$ 0.02 & 1.0 $\pm$ 0.05 \\
Per-emb-12 & 4.7 $\pm$ 0.05 & 6.9 $\pm$ 0.005 & 1.0 $\pm$ 0.01 \\
Per-emb-13 & 2.1 $\pm$ 0.04 & 7.1 $\pm$ 0.009 & 0.89 $\pm$ 0.02 \\
Per-emb-14 & 0.83 $\pm$ 0.03 & 7.9 $\pm$ 0.03 & 1.6 $\pm$ 0.07 \\
Per-emb-15 & 2.3 $\pm$ 0.07 & 6.8 $\pm$ 0.01 & 0.66 $\pm$ 0.02 \\
Per-emb-16 & 0.56 $\pm$ 0.03 & 8.8 $\pm$ 0.03 & 1.2 $\pm$ 0.07 \\
Per-emb-17 & 0.50 $\pm$ 0.03 & 6.0 $\pm$ 0.05 & 1.5 $\pm$ 0.1 \\
Per-emb-18 & 1.00 $\pm$ 0.03 & 8.1 $\pm$ 0.03 & 1.8 $\pm$ 0.08 \\
Per-emb-19 & 1.1 $\pm$ 0.05 & 7.8 $\pm$ 0.02 & 0.99 $\pm$ 0.05 \\
Per-emb-20 & 1.2 $\pm$ 0.04 & 5.3 $\pm$ 0.01 & 0.76 $\pm$ 0.03 \\
Per-emb-21 & 0.48 $\pm$ 0.03 & 9.0 $\pm$ 0.05 & 1.7 $\pm$ 0.1 \\
Per-emb-22 & 2.8 $\pm$ 0.08 & 4.3 $\pm$ 0.02 & 1.2 $\pm$ 0.04 \\
Per-emb-23 & 1.9 $\pm$ 0.05 & 7.8 $\pm$ 0.01 & 0.81 $\pm$ 0.02 \\
Per-emb-24 & 1.0 $\pm$ 0.1 & 7.8 $\pm$ 0.02 & 0.33 $\pm$ 0.04 \\
Per-emb-25 & 0.56 $\pm$ 0.06 & 5.8 $\pm$ 0.04 & 0.71 $\pm$ 0.09 \\
Per-emb-26 & 1.1 $\pm$ 0.02 & 5.4 $\pm$ 0.02 & 1.5 $\pm$ 0.04 \\
Per-emb-27 & 1.4 $\pm$ 0.04 & 8.1 $\pm$ 0.02 & 1.7 $\pm$ 0.05 \\
Per-emb-28 & 0.27 $\pm$ 0.03 & 8.6 $\pm$ 0.05 & 0.93 $\pm$ 0.12 \\
Per-emb-29 & 3.2 $\pm$ 0.08 & 6.4 $\pm$ 0.01 & 0.92 $\pm$ 0.02 \\
Per-emb-30 & 1.6 $\pm$ 0.05 & 6.9 $\pm$ 0.02 & 1.6 $\pm$ 0.05 \\
Per-emb-31 & 0.16 $\pm$ 0.03 & 8.0 $\pm$ 0.2 & 2.6 $\pm$ 0.5 \\
Per-emb-32 & 0.58 $\pm$ 0.06 & 9.4 $\pm$ 0.04 & 0.79 $\pm$ 0.10 \\
Per-emb-33 & 0.70 $\pm$ 0.06 & 5.3 $\pm$ 0.04 & 2.1 $\pm$ 0.09 \\
Per-emb-34 & 0.60 $\pm$ 0.06 & 5.7 $\pm$ 0.05 & 1.7 $\pm$ 0.1 \\
Per-emb-35 & 1.8 $\pm$ 0.05 & 7.4 $\pm$ 0.01 & 1.2 $\pm$ 0.04 \\
Per-emb-36 & 2.3 $\pm$ 0.08 & 6.9 $\pm$ 0.02 & 1.1 $\pm$ 0.04 \\
Per-emb-37 & 1.7 $\pm$ 0.08 & 7.5 $\pm$ 0.009 & 0.41 $\pm$ 0.02 \\
Per-emb-38 & 0.59 $\pm$ 0.05 & 7.1 $\pm$ 0.04 & 0.92 $\pm$ 0.09 \\
Per-emb-40 & 1.2 $\pm$ 0.06 & 7.4 $\pm$ 0.05 & 2.1 $\pm$ 0.1 \\
Per-emb-41 & 0.38 $\pm$ 0.08 & 6.5 $\pm$ 0.07 & 0.73 $\pm$ 0.18 \\
Per-emb-42 & 0.62 $\pm$ 0.04 & 5.8 $\pm$ 0.03 & 0.94 $\pm$ 0.06 \\
Per-emb-44 & 2.8 $\pm$ 0.05 & 8.7 $\pm$ 0.02 & 2.0 $\pm$ 0.04 \\
Per-emb-46 & 0.28 $\pm$ 0.06 & 5.2 $\pm$ 0.09 & 0.93 $\pm$ 0.22 \\
Per-emb-47 & 0.34 $\pm$ 0.03 & 7.4 $\pm$ 0.04 & 1.0 $\pm$ 0.09 \\
Per-emb-48 & 0.24 $\pm$ 0.03 & 4.7 $\pm$ 0.06 & 0.88 $\pm$ 0.15 \\
Per-emb-50 & 0.25 $\pm$ 0.03 & 7.6 $\pm$ 0.1 & 1.8 $\pm$ 0.3 \\
Per-emb-51 & 0.27 $\pm$ 0.04 & 7.0 $\pm$ 0.04 & 0.62 $\pm$ 0.10 \\
Per-emb-52 & 0.34 $\pm$ 0.06 & 8.3 $\pm$ 0.03 & 0.43 $\pm$ 0.08 \\
Per-emb-53 & 1.2 $\pm$ 0.04 & 11.0 $\pm$ 0.02 & 1.3 $\pm$ 0.05 \\
Per-emb-54 & 2.8 $\pm$ 0.04 & 7.9 $\pm$ 0.009 & 1.4 $\pm$ 0.02 \\
Per-emb-55 & 0.46 $\pm$ 0.05 & 12.0 $\pm$ 0.05 & 1.0 $\pm$ 0.1 \\
Per-emb-56 & 0.64 $\pm$ 0.12 & 11.0 $\pm$ 0.02 & 0.31 $\pm$ 0.08 \\
Per-emb-58 & 1.7 $\pm$ 0.09 & 8.3 $\pm$ 0.02 & 0.63 $\pm$ 0.04 \\
Per-emb-61 & 0.33 $\pm$ 0.04 & 9.3 $\pm$ 0.07 & 1.0 $\pm$ 0.2 \\
Per-emb-62 & 0.74 $\pm$ 0.08 & 8.6 $\pm$ 0.02 & 0.41 $\pm$ 0.05 \\
B1-bN & 0.16 $\pm$ 0.05 & 8.0 $\pm$ 0.3 & 1.7 $\pm$ 0.6 \\
Per-bolo-58 & 0.44 $\pm$ 0.05 & 8.2 $\pm$ 0.03 & 0.59 $\pm$ 0.08 \\
SVS 13B & 1.9 $\pm$ 0.07 & 8.5 $\pm$ 0.01 & 0.68 $\pm$ 0.03 \\
SVS 13C & 2.1 $\pm$ 0.05 & 8.9 $\pm$ 0.02 & 1.9 $\pm$ 0.05 \\
\enddata
\tablecomments{Fits are for a single-Gaussian component only, even if the spectra have features of multiple components. The reported uncertainties in the table are the fitting uncertainties only.}
%\tablecomments{Although \coto, \ttco, and \ceo\ are essentially detected toward every pointing, it does not mean that the line is associated with the protostar. Large-scale emission from the Perseus molecular cloud is detected with the SMA even when emission is not associated with the protostar.}
\tablenotetext{a}{FWHM linewidths measured across the single-component Gaussian fit.}
%\tablenotetext{b}{These contours are those shown in Figure~\ref{outflows}.}
%\tablenotetext{b}{RA and DEC are given for the phase center of the observations.}
%\tablenotetext{c}{These tracks were missing the ASIC chunks for \coto\ and \ttco.}
%\tablenotetext{*}{These spectral lines that were not used in this study.} 
\end{deluxetable}

\section{Analysis of Select Targets}\label{sec:select}

In this section, we give a brief overview of many candidate protostars in the MASSES survey. We specifically question the candidacy of certain protostars, we provide a brief analysis of the first hydrostatic core candidates, and we analyze the SVS~13 system, which was observed with the SWARM correlator.

\subsection{Questionable Protostars in the MASSES Sample}
We noted in \citetalias{Stephens2018} that the protostellar classification for 6 MASSES targets is questionable. These targets had no spectral line or continuum emission obviously associated with a protostar. They were also undetected in the VANDAM survey \citep{Tobin2016}. Moreover, they generally had poor constraints on their bolometric luminosities, with the derived luminosities being consistent with 0 \citep{Enoch2009}. These protostars are Per-emb-4, Per-emb-39, Per-emb-43, Per-emb-45, Per-emb-59, and Per-emb-60. The observations in this release each have SWARM EXT data added to the previous observations, which allows for a substantial improvement to the continuum sensitivity over \citetalias{Stephens2018}. Based on Equation~\ref{m_disk_eq_limit} and the sensitivities listed in Table~\ref{tab:sens}, the 1.3\,mm mass sensitivity is about 0.0075~$M_\odot$, except for Per-emb-39, which is about 0.010~$M_\odot$. The sensitivity at 850~$\mu$m is notably worse given the lack of SWARM EXT tracks toward these protostars, although Per-emb-60 has a marginal detection at 850\,$\mu$m. Continuum is again not detected toward any of these protostars, except for Per-emb-39. However, Per-emb-39 is very extended, and the continuum is possibly an artifact from resolving out large-scale emission. 

In \citetalias{Stephens2018}, we also noted that compact emission is detected with ground-based single-dish observations and/or $Herschel$ toward Per-emb-4, Per-emb-39, and Per-emb-60, but not toward Per-emb-43, Per-emb-45, and Per-emb-59. Thus, these latter three protostellar candidates are even less likely to be protostars. Overall, based on the improved continuum sensitivity and the lack of line emission, we conclude that all six of these sources are unlikely to be protostars.
%Per-emb-4    0.47
%Per-emb-39  0.60
%Per-emb-43  0.42
%Per-emb-45  0.47
%Per-emb-59  0.46
%Per-emb-60  0.47

\subsection{Candidate First Cores}\label{sec:cfc}
\citet{Larson1969} first suggested the presence of the first hydrostatic cores, where collapsing material onto a central dense object first approaches hydrostatic equilibrium. The true protostar does not form until more mass is accreted and the central temperature increases to 2000~K, where molecular hydrogen dissociates, causing dynamical collapse. The first hydrostatic core period is expected to be short-lived, and have low-velocity, uncollimated outflows \citep[e.g.,][]{Machida2008,Matsumoto2011,Tomisaka2011,Machida2014}. First hydrostatic cores were not identified by the shallow $Spitzer$ IRAC surveys \citep[e.g.,][]{Evans2003,Evans2009,Enoch2009}, but could be identified with (sub)millimeter and/or mid-IR observations \citep[e.g.,][]{Enoch2006,Enoch2010}.

The MASSES survey observed six first-core candidates, all of which have been studied before. These sources are B1-bN and B1-bS \citep[e.g.,][]{Pezzuto2012}, L1448IRS2e \citep[e.g.,][]{ChenX2010}, L1451-MMS \citep[henceforth, L1451-mm; e.g.,][]{Pineda2011,Maureira2017a}, Per-bolo-45 \citep[e.g.,][]{Schnee2012}, and Per-bolo-58 \citep[e.g.,][]{Enoch2010, Dunham2011,Maureira2017b}. For four of these first-core candidates (B1-bN, B1-bS, L1451-mm, and Per-bolo-58), both the MASSES and VANDAM surveys detected compact continuum data (e.g., Figure~\ref{fig:contn2dp}). These same four objects all appear to have outflowing gas. The outflowing gas for Per-bolo-58 and L1451-mm is particularly slow, only extending up to $\sim$4 and 1\,\kms\ from the systemic velocity, respectively. The outflow for L1451-mm is particularly compact, with a dynamical time of only a few thousand years \citep{Pineda2011}, consistent with the lifetime of a first hydrostatic core \citep{Machida2008}. B1-bN and B1-bS outflows are slightly faster, reaching velocities of up to $\sim$10\,\kms\ from systemic. Of these four first-core candidates, L1451-mm is perhaps the most promising candidate given its very compact and slow outflow.

The other two candidates, L1448~IRS2E and Per-bolo-45, were not detected by the VANDAM survey and are more ambiguous. They are discussed in more detail below.

\subsubsection{L1448 IRS2E}
\citet{OLinger1999} first identified the L1448~IRS2E core with the James Clerk Maxwell Telescope (JCMT) telescope using the SCUBA camera at 450 and 850\,$\mu$m. \citet{ChenX2010} followed up with 1.3\,mm SMA observations to determine whether it is a promising first hydrostatic core candidate. With a sensitivity of  0.85\,mJy\,bm$^{-1}$ at a resolution of 3$\farcs$9~$\times$~2$\farcs$6, they found an unresolved continuum source with an integrated flux of approximately 6~$\pm$~2\,mJy, corresponding to $\sim$0.04\,$M_\odot$. This continuum source was undetected in \citetalias{Stephens2018}, but the sensitivities of these observations were only 2.7\,mJy\,bm$^{-1}$. As such, we could not significantly call this a nondetection.

In this MASSES data release, we combine the previous L1448 IRS2E data with an EXT SWARM track to improve the sensitivity to 0.79\,mJy\,bm$^{-1}$ at a resolution of 1$\farcs$1~$\times$~0$\farcs$9 (Table~\ref{tab:sens}). Because the source was unresolved for \citet{ChenX2010}, we would expect to easily detect a 6~$\pm$~2\,mJy source with these observations. However, no continuum source is detected by the MASSES survey at 1.3\,mm (Figure~\ref{fig:rp1outflows}). No source was detected at 850~$\mu$m either, though the single MASSES subcompact track had a sensitivity of only 10.0\,mJy\,bm$^{-1}$. Nevertheless, the source certainly could be marginally detected at 850~$\mu$m, as a 6\,mJy source would be approximately 20--30\,mJy at 850~$\mu$m, assuming a $\beta$ between 1 and 2. Based on Equations~\ref{m_disk_eq_limit} and~\ref{m_disk_eq_limit850}, the 3$\sigma$ upper limit for a compact source within the beam is 0.013 and 0.048~$M_\odot$ at 1.3\,mm and 850\,$\mu$m, respectively. A 1.3\,mm, 6\,mJy source, however, would have a mass of 0.032\,$M_\odot$ using the assumptions in Section~\ref{sec:mass_sensitivity}.

%0.79 and 10
These MASSES observations question the existence of a compact continuum source of 6\,mJy, unless the source was flaring in the past due to an accretion burst. Nevertheless, some evidence of a young protostar exists as there appears to be a high-velocity red outflow lobe (Figure~\ref{fig:rp1outflows}). While the observed red-shifted emission from the mostly west--east direction is from an outflow cavity wall from Per-emb-22 (L1448~IRS2), the red emission also extending toward the north--south direction appears to be outflowing gas, as the CO emission reaches velocities of over 25\,\kms\ from the systemic velocity. No other source in Perseus \citep[e.g., those defined in][]{Young2015} appears to drive this outflowing gas. The high-velocity outflowing gas, however, is expected to be inconsistent with what would be expected for a first hydrostatic core \citep[e.g.,][]{Machida2008}. Based on these models, if L1448 IRS2E is a true young stellar object, it is unlikely to be a first hydrostatic core. Moreover, the north--south \coto\ emission in the L1448~IRS2E pointing is coincident along the line of sight with Per-emb-22's outflow cavity, suggesting that such emission may be caused by a shock resulting from the deflection of the outflow from IRS2 interacting with the surrounding dense cloud material \citep[e.g.,][]{Kajdic2012}. 
%6(1.3/0.87)^3 = 20 mJy/bm

\subsubsection{Per-bolo-45}
Per-bolo-45 is located in NGC~1333 and is perhaps the least studied of the six first core candidates. It was identified as a starless core candidate in \citet{Sadavoy2010a}. \citet{Schnee2012} used the Combined Array for Research in Millimeter-wave Astronomy (CARMA) at 3\,mm and found an extended continuum source located northwest from the phase center, in the same location in which we also detect a 1.3\,mm continuum source (Figure~\ref{fig:contn2dp}). The detected 1.3\,mm continuum source is contained within a noisy image, and the noise seems in part due to the SMA resolving out emission. The continuum emission detected by CARMA and the SMA is notably extended, and no VANDAM source was detected toward this pointing (Figure~\ref{fig:contn2dp}). \citet{Schnee2012} detected \mbox{SiO(2--1)} emission at 8 to 9\,\kms\ significantly south of the phase center (i.e., away from the millimeter continuum source). We also detect both \coto\ and \cott\ emission in a single 0.5\,\kms\ channel centered at 7.5\,\kms\ that is broadly consistent with this \mbox{SiO(2--1)} emission. We also detect SiO(5--4) for a couple of channels around 8\,\kms\ at the same position as the SiO(2--1). The origin of the \mbox{SiO(2--1)}, \coto, and \cott\ emission at $\sim$7.5\,\kms\ is a bit ambiguous, and given the kinematics of NGC~1333 and the fact that there are other young stellar objects nearby, such as Class~II sources \citep{Young2015}, we can only consider this as marginal evidence of a low-velocity outflow from Per-bolo-45. We also note that there is higher velocity \coto, \cott\ (Figure~\ref{fig:rp1outflows} and~\ref{fig:co32outflows}), and \mbox{CO(1--0)} \citep{Plunkett2013} emission detected west of the continuum source, but given the orientation of the emission, it is not likely ejected from the protostar.

 %Since this emission is only for one channel and is at the systemic velocity of the cloud (i.e., the interferometer is confused with the large-scale emission at these velocities), we cannot confirm that this emission is truly an outflow. 

\citet{Schnee2012} found many other spectral lines likely associated with the millimeter continuum source for Per-bolo-45, including \mbox{NH$_2$D(1$_{1,1}$--1$_{0,1}$)}, \mbox{HCO$^+$(1--0)}, \mbox{HNC(1--0)}, and \mbox{N$_2$H$^+$(1--0)}. The spectral line \mbox{NH$_2$D(1$_{1,1}$--1$_{0,1}$)} is particularly correlated with the continuum source. Similarly, Figure~\ref{fig:contn2dp} shows that \ntdp\ is strongly correlated with the 1.3\,mm continuum source. Moreover, even though \citet{Schnee2012} found \mbox{HCO$^+$(1--0)} slightly offset from the 3\,mm continuum peak, we did not detect \hcop\ throughout the entire map, which may be due to the fact that it is a higher excitation line. Given that these two deuterated species are detected toward the continuum, \mbox{HCO$^+$(1--0)} is offset from the continuum (i.e., CO is frozen to the grains at the peak), and the higher excitation line \hcop\ is not detected, Per-bolo-45 indeed appears to be a very cold source.

First cores can have a variety of temperatures and radii, which largely depend on the initial conditions of the collapse and the age of the first core. For example, \citet{Bhandare2018} showed that radii can vary from $\sim$1 to 10\,au, and the first-core temperatures start at $\sim$10~K and rise to $\sim$2000~K before collapsing to a second (i.e., protostellar) core. The VANDAM survey observed Per-bolo-45 with the VLA in the B configuration only and had a sensitivity of $\sim$0.1\,K at 8\,mm. Thus, the VANDAM survey is expected to detect the more evolved and hotter first cores, but would not detect the youngest or smallest first cores.

Based on the above, Per-bolo-45 is probably not protostellar due to the lack of a compact source or well-defined outflow, and it is a cold source. It could be a starless core \citep[as originally identified by][]{Sadavoy2010a} with a significant concentration peak that can be detected by interferometers with sufficiently short baselines. Based on deeper ALMA observations, substructure is only expected toward starless cores that are gravitationally unstable \citep{Kirk2017}, as the ones stable to collapse show no substructure \citep{Dunham2016}. We suggest that Per-bolo-45 is at a rare evolutionary stage, with features indicating it is a starless core on the verge of collapse. Nevertheless, we certainly cannot rule out that it is a first core due to the detected extended continuum and the marginal evidence of an outflow. In the case it is a first core, it is likely at the earliest stage when the first core is expected to be cold.
 
%However, \citet{Dunham2016} observed 56 starless cores in Chamaeleon with 2.8\,mm ALMA continuum observations, and detected no sources; a source such as Per-bolo-45 would have easily been detected in that sample.

\begin{figure*}[ht!]
\begin{center}
\includegraphics[width=2\columnwidth]{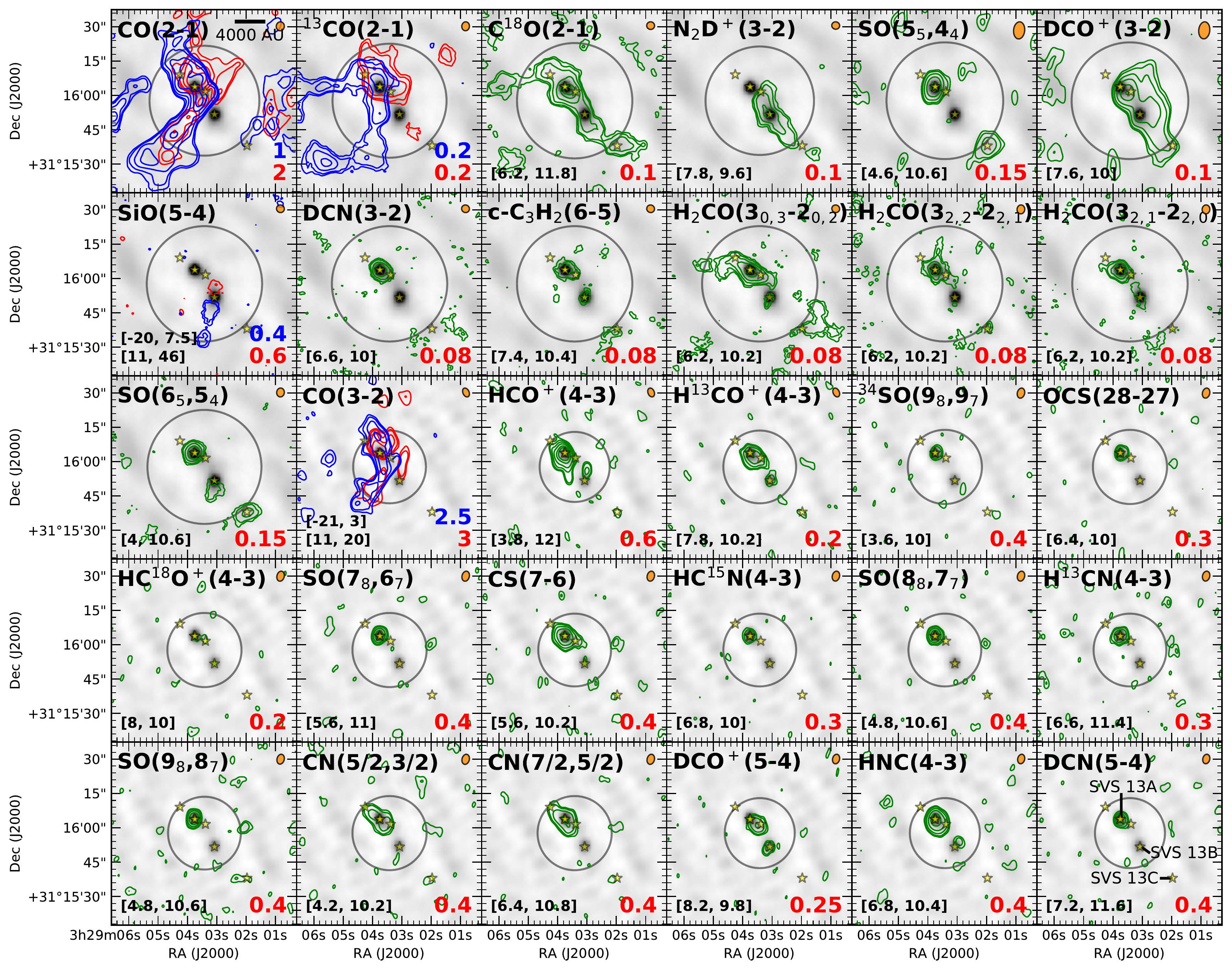}
\end{center}
\caption{Spectral lines detected with the SWARM correlator toward the pointing containing Per-emb-44 (SVS~13A) and SVS~13B, using a weighting of robust~=~--1. The locations of the main SVS continuum sources are indicated in the bottom-right panel. The 1.3\,mm or 850~$\mu$m continuum is in grayscale, depending on the receiver used for the particular spectral line. The contours show the integrated intensity, where \coto, \ttco, \mbox{SiO(5--4)}, and \cott\ are separated into blue and red components, while others are integrated over a single velocity range. The integrated velocity ranges are indicated in brackets at the bottom left of each panel where [start,~stop] indicate the start and stop of the velocity integration range in units of \kms. For the unlabeled panels, \coto\ ranges are [--17,~3] (blue) and [11,~20] (red) and \ttco\ are [0.5,~6.8] (blue) and [10.7,~14] (red). Stars indicate the location of known protostars identified in \citet{Tobin2016}. The synthesized beam for the spectral line is indicated by the orange ellipse in the top-right of each panel. Gray circles indicate the FWHM of the primary beam. Images have not been corrected for the primary beam.
}%Note that the majority of the pointings use the ASIC correlator and the SUB configuration, which do not map all these lines. 
\label{fig:Per44} 
\end{figure*}

\subsection{The SVS~13 Star-Forming Region}
\begin{figure*}[ht!]
\begin{center}
\includegraphics[width=2\columnwidth]{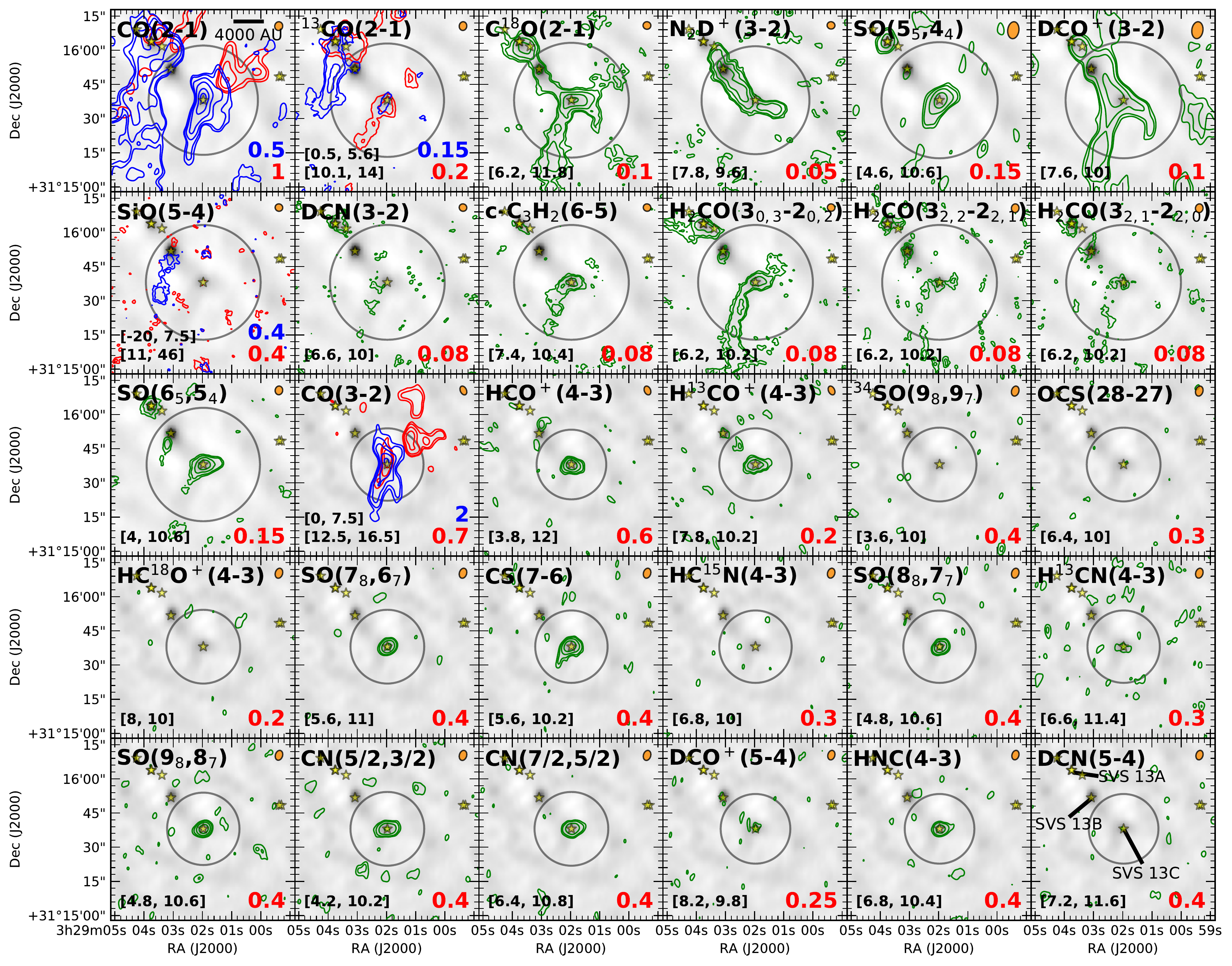}
\end{center}
\caption{Same as Figure~\ref{fig:Per44}, except now the image is centered on the SVS~13C pointing. The integrated velocity ranges for \coto\ are [0,~7] (blue) and [12.5,~16.5] (red), and the \ttco\ ranges are indicated in its panel. 
}
\label{fig:SVS13C} 
\end{figure*}

Of the 18 SUB pointings that contained SWARM data, the Per-emb-44/SVS~13B pointing has the most spectral lines detected across the entire SWARM bandwidth. The SVS~13 system \citep{Strom1976} contains three main continuum sources, which, from northeast to southwest (Figure~\ref{fig:Per44} and~\ref{fig:SVS13C}), are called SVS~13A (Per-emb-44 in \citealt{Enoch2009}), SVS~13B, and SVS~13C \citep{Looney2000}. At 32.5\,$L_\odot$, SVS~13A has the highest estimated bolometric luminosity of all protostars in the MASSES sample. We have two pointings toward this system, with one pointing centered between SVS~13A and SVS~13B and the other centered on SVS~13C, and images of each spectral line are shown in Figure~\ref{fig:Per44} and~\ref{fig:SVS13C}, respectively. Every spectral line listed in Table~\ref{tab:lines} and~\ref{tab:swarmlines} was detected using the SWARM correlator toward the SVS system. 

SVS~13A is likely the most evolved source, as it has the largest $T_{\text{bol}}$. Toward its continuum peak, all lines are detected except for \mbox{SiO(5--4)} and \ntdp. \ntdp\ peaks at the continuum position of SVS~13B and is also detected at SVS~13C, although the line emission has a local minima at the continuum peak of that protostar. N$_2$D$^+$ is expected to disappear at temperatures $>$20~K \citep[e.g.,][]{Jorgensen2011}, and given the high temperature of SVS~13A, the nondetection is expected \citep[also see, e.g.,][]{Emprechtinger2009,Tobin2013c}. The fact that N$_2$D$^+$ peaks for SVS~13B and has a local minima for SVS~13C may imply that SVS~13B is colder and perhaps less evolved than SVS~13C. N$_2$D$^+$ may be absent near the SVS~13C protostar where it is much hotter, and the outer envelope of SVS~13C is still cold enough for there to be some N$_2$D$^+$ along the line of sight. This line of reasoning would also suggest that the outer envelope of SVS~13A is much hotter than that of 13B and 13C.

Besides N$_2$D$^+$, we also observed two other deuterated species, i.e., DCO$^+$ (both \mbox{3--2} and \mbox{5--4}) and DCN (also both \mbox{3--2} and \mbox{5--4}). DCO$^+$ is strongly detected toward all three sources, while DCN is only detected toward SVS~13A (though DCN toward SVS~13C is marginal). N$_2$D$^+$ is only formed via low-temperature pathways, while DCN and DCO$^+$ have both low and high-temperature pathways \citep[e.g., see][for discussion]{Salinas2017}, though the DCO$^+$ high-temperature pathway is not likely on envelope scales. For DCN, the high temperature pathway is expected to be the dominant formation mechanism \citep[66\%;][]{Turner2001}. DCO$^+$ in general is more abundant than DCN and N$_2$D$^+$, and is not as readily destroyed at high temperatures, so it makes sense to see it detected across the entire SVS~13 system. DCN also can be formed at low and high temperatures, and the lack of detection toward SVS~13B and 13C may simply be due to the fact that we lack the sensitivity to detect the molecule, especially given that a similar molecule, H$^{13}$CN, is weak or undetected toward these two protostars. Nevertheless, DCN is typically more likely to be detected toward warmer sources than cold sources \citep{Jorgensen2004}.

The continuum emission for SVS~13B is substantially brighter than that for SVS~13C, and its envelope mass is estimated to be $\sim$4 times larger \citepalias{Stephens2018}. However, most spectral lines (e.g., SO and CN) are significantly brighter toward SVS~13C. Again, this supports the idea that SVS~13C is warmer and perhaps more evolved than SVS~13B.

The only other line not detected toward SVS~13A is SiO(5--4), which is detected toward SVS~13B over a wide velocity range. SiO is typically associated with shocks because shocks can release silicon from dust grains \citep[e.g.,][]{Schilke1997}. The emission is morphologically consistent with a highly collimated bipolar outflow. This outflow was originally identified via SiO(2--1) observations by \cite{Bachiller1998,Bachiller2000}.

\section{Summary}
MASSES is an unbiased protostellar survey that used the SMA to target continuum and spectral lines toward all 74 known Class~0 and~I protostellar candidates in the Perseus molecular cloud. Using observations at both 1.3\,mm and 850\,$\mu$m, we observed the continuum at both wavelengths, as well as \cott, \ttco, \ceo, \ntdp, \cott, \hcop, and \httcop. Additional spectral lines were also observed toward 18 targets, as midway through the MASSES survey, the correlator was upgraded from ASIC to SWARM, which provides high spectral resolution throughout its entire bandwidth. All 1.3\,mm targets were observed in both the SUB and EXT SMA configurations, while 850\,$\mu$m was only observed in the SUB configuration. The first MASSES data release, \citetalias{Stephens2018}, only released SUB data at 1.3\,mm and did not include any additional lines detected with SWARM. This paper releases visibilities and imaged data for the full SUB+EXT configuration (and COM, when available), including additional lines detected with SWARM. We also provide images/data cubes for two different interferometric weightings, with resolutions as high as $\sim$1$\arcsec$ ($\sim$300\,au). The data are publicly available at \url{https://dataverse.harvard.edu/dataverse/full_MASSES/}.

As compared to the subcompact data release \citepalias{Stephens2018}, this data release has a substantial improvement in data products. Firstly, the SMA recently found an error with how they been doing the Doppler tracking over the past several years (see Section~\ref{sec:datacal}), which has been fixed for this data release. We also slightly improved the imaging by removing ``options=positive" during \texttt{clean}, as discussed in Section~\ref{sec:imaging}. As such, it is suggested that the user use these data release products over the SUB products from \citetalias{Stephens2018}.

We give a survey overview and find the following results:
\begin{enumerate}
\item For the continuum and the main spectral line tracers (Table~\ref{tab:lines}), 84\% of the delivered images and cubes have significant emission. 
\item \ntdp\ and \httcop\ are more likely to be detected toward younger protostars.
\item Distinct red/blue outflow lobes are more typically associated with younger protostars. %Distinct red/blue outflow lobes are more typically associated with younger protostars
\item Median \ceo\ FWHM linewidths measured toward protostars are $\sim$1.0\,\kms, which are slightly larger than \ceont\ linewidths measured toward other targets via single-dish telescopes. Thus, the \ceo\ linewidths measured about envelopes may be affected by outflows, infall, and/or rotation.
\item As with \citetalias{Stephens2018}, we question the protostellar nature of six targets (Per-emb-4, Per-emb-39, Per-emb-43, Per-emb-45, Per-emb-59, and Per-emb-60), and we constrain their envelope plus disk masses to be $<$0.01\,$M_\odot$.
\item Of the six observed first-core candidates, L1451-mm appears to be the most likely first-core candidate. We argue that Per-bolo-45 may be a starless core on the verge on the verge of collapsing, but we cannot rule out the possibility that it is a first core.
\item Given the improved sensitivity of our observations, it is unlikely that L1448~IRS2E has a compact protostellar source that is 6\,mJy, as originally reported in \citet{ChenX2010}. However, there certainly could be protostellar activity or a shock at its location, given that red-shifted emission is detected with velocities of $\sim$25\,\kms\ above the systemic velocity.
\item Over 30 unique spectral line transitions are detected toward the SVS~13 system. Lines are the brightest toward SVS~13A and the weakest toward 13B. \ntdp\ is the exception, with the opposite trend (i.e., brightest toward 13B and weakest toward 13A). We suggest that SVS~13A is the most evolved, SVS~13B is the least evolved, and SVS~13C is at an evolutionary stage is between the two.
\end{enumerate}

MASSES is the largest publicly available interferometric survey that combines both continuum observations and spectral line observations of protostars.  This survey provides a wealth of information, including envelope structure, outflow morphologies, and early-time chemistry for a complete study of all protostars within in a single cloud.

\acknowledgements
I.W.S. acknowledges support from NASA grant NNX14AG96G. 
T.L.B. acknowledges support from the SMA and the Center for Astrophysics $|$ Harvard \& Smithsonian's Radio and Geoastronomy Division.
E.I.V. acknowledges support from the Russian Ministry of Education and Science grant 3.5602.2017.
J.K.J. acknowledges support from the European Research Council (ERC) under the European Union's Horizon 2020 research and innovation programme (grant agreement No~646908).
%J.J.T. acknowledges support from the University of Oklahoma, the Homer L. Dodge endowed chair, and grant 639.041.439 from the Netherlands Organisation for Scientific Research (NWO).  
%J.E.P. acknowledges the financial support of the European Research Council (ERC; project PALs 320620).
The authors thank the SMA staff for executing these observations as part of the queue schedule; Charlie Qi, Glen Petipas, Qizhou Zhang, and Garrett `Karto' Keating for their technical assistance with the SMA data; and Eric Keto for his guidance with SMA large-scale projects.
We thank SMA director Raymond Blundell for his continued support of the MASSES project.
We thank Manuel Fern\'andez-L\'opez and Robert Sault for useful feedback on MIRIAD imaging. 
The Submillimeter Array is a joint project between the Smithsonian Astrophysical Observatory and the Academia Sinica Institute of Astronomy and Astrophysics and is funded by the Smithsonian Institution and the Academia Sinica.
%This work is based primarily on observations obtained with the SMA, a joint project between the Smithsonian Astrophysical Observatory and the Academia Sinica Institute of Astronomy and Astrophysics and funded by the Smithsonian Institution and the Academia Sinica.
%This research has made use of the VizieR catalogue access tool and the SIMBAD database operated at CDS, Strasbourg, France. 
%This research made use of APLpy \citep{Robitaille2012} and PySpecKit \citep{Ginsburg2011}, which are open-source plotting packages for Python.

\facilities{SMA.}
\software{SMA-MIR (\url{https://github.com/qi-molecules/sma-mir}), MIRIAD \citep{Sault1995}, APLpy \citep{Robitaille2012}, PySpecKit \citep{Ginsburg2011}.}

\appendix
\setcounter{table}{0}
\renewcommand{\thetable}{A\arabic{table}}

\section{MASSES Tables}
This appendix includes four tables, which are referred to in the main text of the paper. Table~\ref{tab:sens} and \ref{tab:sens850} give the beam sizes and sensitivities of all MASSES images for 1.3\,mm and 850\,$\mu$m, respectively. Table~\ref{tab:detection} indicates which tracers were detected toward each protostar. Table~\ref{tab:contours} lists the contours and integrated velocity ranges used for Figures~\ref{fig:rp1outflows}, \ref{fig:rp-1outflows}, and~\ref{fig:co32outflows}.  The contour levels are given by [start, step], where `start' indicates the first contour level, and `step' indicates the increase for each additional contour level. 

\startlongtable
\renewcommand{\tabcolsep}{0.05cm}

\end{longrotatetable}

\bibliography{stephens_bib}
%\bibliography{/Users/istephens/Documents/latex_stuff/stephens_bib}

\end{document}